\documentclass[longauth]{aa}

\usepackage{txfonts}
\usepackage[colorlinks=true,citecolor=blue,linkcolor=blue]{hyperref}

\usepackage[utf8]{inputenc}
\usepackage{graphicx}
\usepackage{amsmath}
\usepackage{amssymb} % Extra maths symbols
\usepackage{enumitem} % customized lists
\usepackage{pifont}% http://ctan.org/pkg/pifont
\newcommand{\cmark}{\ding{51}}%
\newcommand{\xmark}{\ding{55}}%

\usepackage{longtable}
\usepackage{xtab,booktabs}

\newcommand{\snia}{SN~Ia}
\newcommand{\sneia}{SNe~Ia}

\let\ts=\thinspace
\newcommand{\one}{\ts {\sc i}}
\newcommand{\two}{\ts {\sc ii}}
\newcommand{\three}{\ts {\sc iii}}
\newcommand{\four}{\ts {\sc iv}}
\newcommand{\five}{\ts {\sc v}}
\newcommand{\six}{\ts {\sc vi}}

\renewcommand\ion[2]{#1\,\,{\sc{\romannumeral #2}}}

\newcommand{\nifs}{\ensuremath{^{56}\rm{Ni}}}

\newcommand{\cofs}{\ensuremath{^{56}\rm{Co}}}

\newcommand{\fefs}{\ensuremath{^{56}\rm{Fe}}}

\newcommand{\msun}{\ensuremath{\rm{M}_{\odot}}}

\newcommand{\mtot}{\ensuremath{M_\mathrm{tot}}}

\newcommand{\kms}{\ensuremath{\rm{km\,s}^{-1}}}
\newcommand{\gcc}{\ensuremath{\rm{g\,cm}^{-3}}}

\newcommand{\mch}{\ensuremath{M_{\rm Ch}}}

\newcommand{\sr}{\scriptsize}

\usepackage{color}

\begin{document} 

    \title{StaNdaRT: A repository of standardised test models and outputs for supernova radiative transfer}
    \titlerunning{Standardised test models and outputs for SN radiative transfer}

   \author{
     Stéphane~Blondin~(CMFGEN)\inst{1,2,27} %ORCID:  0000-0002-9388-2932
     \and
     Sergei~Blinnikov~(STELLA)\inst{3,4,5,31} %ORCID: 0000-0002-5726-538X
     \and
     Fionntan~P.~Callan~(ARTIS)\inst{6,26} %ORCID: :0000-0002-7975-8185
     \and
     Christine~E.~Collins~(ARTIS)\inst{6,7,26} % ORCID: 0000-0002-0313-7817
     \and
     Luc~Dessart~(CMFGEN)\inst{8,27} %ORCID: 0000-0003-0599-8407
     \and
     Wesley~Even~(SuperNu)\inst{9,10,33} %ORCID: 0000-0002-5412-3618
     \and
     Andreas~Fl\"ors~(TARDIS)\inst{7,34} %ORCID: 0000-0003-2024-2819
     \and
     Andrew~G.~Fullard~(TARDIS)\inst{11,34} %ORCID: 0000-0001-7343-1678
     \and
     D.~John~Hillier~(CMFGEN)\inst{12,27} %ORCID 0000-0001-5094-8017
     \and
     Anders~Jerkstrand~(SUMO)\inst{13,32} % ORCID 0000-0001-8005-4030
     \and
     Daniel~Kasen~(SEDONA)\inst{14,30} % ORCID: 0000-0002-5981-1022
     \and
     Boaz~Katz~(URILIGHT)\inst{15,35} %ORCID: 0000-0003-0584-2920
     \and
     Wolfgang~Kerzendorf~(TARDIS)\inst{11,16,34} %ORCID: 0000-0002-0479-7235
     \and
     Alexandra~Kozyreva~(STELLA)\inst{17,31} %ORCID: 0000-0001-9598-8821
     \and
     Jack~O'Brien~(TARDIS)\inst{11,34} %ORCID: 0000-0003-3615-9593
     \and
     Ezequiel~A.~Pássaro~(TARDIS)\inst{18,34} %ORCID: 0000-0002-3707-1975
     \and
     Nathaniel~Roth~(SEDONA)\inst{30} %ORCID 0000-0002-6485-2259
     \and
     Ken~J.~Shen~(SEDONA)\inst{19,30} %ORCID 0000-0002-9632-6106
     \and
     Luke~Shingles~(ARTIS)\inst{6,7,26} %ORCID: 0000-0002-5738-1612
     \and
     Stuart~A.~Sim~(ARTIS)\inst{6,26} %ORCID: 0000-0002-9774-1192
     \and
     Jaladh~Singhal~(TARDIS)\inst{34} %ORCID: 0000-0002-8310-0829
     \and
     Isaac~G.~Smith~(TARDIS)\inst{11,34} %ORCID: 0000-0003-0440-3918
     \and
     Elena~Sorokina~(STELLA)\inst{3,4,31} %ORCID: N/A
     \and
     Victor~P.~Utrobin~(CRAB)\inst{3,20,28} % ORCID: 0000-0001-7056-4571
     \and
     Christian~Vogl~(TARDIS)\inst{17,33} %ORCID: 0000-0002-7941-5692
     \and
     Marc~Williamson~(TARDIS)\inst{21,34} %ORCID: 0000-0003-2544-4516
     \and
     Ryan~Wollaeger~(SuperNu)\inst{22,23,33} % ORCID: 0000-0003-3265-4079
     \and
     Stan~E.~Woosley~(KEPLER)\inst{24,29} %ORCID: 0000-0002-3352-7437
     \and
     Nahliel~Wygoda~(URILIGHT)\inst{25,35} %ORCID: N/A
   }
   \authorrunning{S. Blondin et al.}

   \institute{
     Aix Marseille Univ, CNRS, CNES, LAM, Marseille, France\\
     \email{stephane.blondin@lam.fr}
     \and     
     Unidad Mixta Internacional Franco-Chilena de Astronom\'ia, CNRS/INSU UMI 3386 and Instituto de Astrof\'isica,\\
     Pontificia Universidad Cat\'olica de Chile, Santiago, Chile
     \and
     NRC ``Kurchatov institute'', Institute for Theoretical and Experimental Physics (ITEP), Moscow 117218, Russia
     \and
     Sternberg Astronomical Institute, Moscow State University, Moscow 119234, Russia
     \and
     Kavli Institute for the Physics and Mathematics of the Universe (WPI), The University of Tokyo Institutes for Advanced Study,\\
     The University of Tokyo, 5-1-5 Kashiwanoha, Kashiwa, Chiba 277-8583, Japan
     \and
     Astrophysics Research Centre, School of Mathematics and Physics, Queen's University Belfast, Belfast BT7 1NN, Northern Ireland, UK
     \and
     GSI Helmholtzzentrum f\"ur Schwerionenforschung, Planckstraße 1, 64291 Darmstadt, Germany
     \and
     Institut d'Astrophysique de Paris, CNRS-Sorbonne Université, 98 bis boulevard Arago, 75014, Paris, France
     \and
     Center for Theoretical Astrophysics, Los Alamos National Laboratory, Los Alamos, NM, 87545, USA
     \and
     Theoretical Division, Los Alamos National Laboratory, Los Alamos, NM, 87545, USA
     \and
     Department of Physics and Astronomy, Michigan State University, East Lansing, MI 48823, USA
     \and
     Department of Physics and Astronomy \& Pittsburgh Particle Physics, Astrophysics, and Cosmology Center (PITT PACC),\\
     University of Pittsburgh, 3941 O’Hara Street, Pittsburgh, PA 15260, USA
     \and
     The Oskar Klein Centre, Department of Astronomy, Stockholm University, AlbaNova, SE-10691 Stockholm, Sweden
     \and
     Departments of Physics and Astronomy, University of California Berkeley and Lawrence Berkeley National Laboratory, USA
     \and
     Dept. of Particle Phys. \& Astrophys., Weizmann Institute of Science, Rehovot 76100, Israel
     \and
     Department of Computational Mathematics, Science, and Engineering, Michigan State University, East Lansing, MI 48824, USA
     \and
     Max-Planck-Institut f\"ur Astrophysik, Karl-Schwarzschild-Straße 1, 85748 Garching bei M\"unchen, Germany
     \and
     Facultad de Ciencias Astronómicas y Geofísicas, Universidad Nacional de La Plata, La Plata, B1900, Argentina
     \and
     Department of Astronomy and Theoretical Astrophysics Center, University of California, Berkeley, CA 94720, USA
     \and
     Institute of Astronomy, Russian Academy of Sciences, Pyatnitskaya St. 48, 119017 Moscow, Russia
     \and
     Department of Physics, New York University, New York, NY, 10003, USA
     \and
     Computer, Computational, and Statistical Sciences Division, Los Alamos National Laboratory, Los Alamos, NM, 87545, USA
     \and
     Department of Physics \& Astronomy, Louisiana State University, Baton Rouge, LA, 70803, USA
     \and
     Department of Astronomy and Astrophysics, University of California, Santa Cruz, CA 95064, USA
     \and
     Department of Physics, NRCN, Beer-Sheva 84190, Israel
     \and
     ARTIS Collaboration
     \and
     CMFGEN Collaboration
     \and
     CRAB Collaboration
     \and
     KEPLER Collaboration 
     \and
     SEDONA Collaboration 
     \and
     STELLA Collaboration 
     \and
     SUMO Collaboration 
     \and
     SuperNu Collaboration 
     \and
     TARDIS Collaboration 
     \and
     URILIGHT Collaboration 
   }

   \date{Received 27 May 2022; accepted 22 September 2022}

\abstract{
We present the first results of a comprehensive supernova (SN)
radiative-transfer (RT) code-comparison initiative (StaNdaRT), where
the emission from the same set of standardised test models is
simulated by currently used RT codes.  We ran a total of ten codes on a set of four benchmark ejecta models of Type Ia SNe. We
consider two sub-Chandrasekhar-mass ($\mtot = 1.0$\,\msun) toy models
with analytic density and composition profiles and two
Chandrasekhar-mass delayed-detonation models that are outcomes of
hydrodynamical simulations. We adopt spherical symmetry for all four
models. The results of the different codes, including the light
curves, spectra, and the evolution of several physical properties as a
function of radius and time are provided in electronic form in a
standard format via a public repository. We also include the detailed
test model profiles and several Python scripts for accessing and
presenting the input and output files. We also provide the code used
to generate the toy models studied here. In this paper, we describe  the test models, radiative-transfer codes, and output formats in
detail,
and provide access to the repository. We present example results of
several key diagnostic features.
}

\keywords{
supernovae: general --
Radiative transfer
} 

\maketitle

%%%%%%%%%%%%%%%%%%%%%%%%%%%%%%%%%%%%%%%
% Introduction
%%%%%%%%%%%%%%%%%%%%%%%%%%%%%%%%%%%%%%%

\section{Introduction}\label{sect:intro}

Accurate radiative-transfer (RT) calculations remain a key challenge
in the study of astronomical transients such as supernovae (SNe). While
advances in computational capabilities and theoretical understanding
have allowed great progress in the ability to simulate radiation
transport, the large number of physical processes involved, in
particular opacity from thousands of atomic transitions with a mixed
absorptive and scattering character, prohibit comprehensive 3D
calculations based on first principles. Several physical
approximations of different forms ---in particular different treatments
of the significant deviations from local thermodynamic equilibrium
(LTE)--- are employed by different RT codes to calculate the properties
of the gas and of the radiation field. Approximate treatment of atomic
physics is also required due to the partially calibrated atomic data.

The back-reaction of radiation on the hydrodynamics provides an
additional challenge, requiring the solution of hydrodynamic equations
coupled to the RT solution. However, in many cases, and in particular for
Type Ia SNe (\sneia) at phases beyond several days, which we focus on here, the radiation carries a negligible fraction of the
energy and the ejecta are freely expanding homologously. The RT
problem in these cases is decoupled from the hydrodynamics problem,
the latter providing the initial ejecta profiles (`ejecta
models' hereafter). The ejecta profiles include density, composition, and
initial temperature as a function of position. The initial time (of
order 1 day) is such that on the one hand it is much larger than the
explosion timescale (of order 1 second) so that the expansion is
nearly homologous and on the other hand sufficiently early such that
radiation has hardly diffused across the ejecta and the only evolution
is due to adiabatic expansion and radioactive decay.

% table generated with read_inputs.py
% placed here to conform with published version
\begin{table*}
\centering
\footnotesize
\caption{Summary of ejecta conditions. The yields for representative species corresponds to the start of the simulations in our model set (2\,d post explosion for the toy models and $\sim$1\,d post explosion for the DDC models). The $^{56}$Ni mass is given prior to any decay.}
\label{tab:models}
\begin{center}
\begin{tabular}{lccccccccc}
\hline
Model & $M_{\rm ej}$ & $E_{\rm kin}$    & M($^{56}$Ni)$_{t=0}$ & M(Fe)   & M(Ca)   & M(S)    & M(Si)   & M(O)     & M(C)    \\
      & (\msun)      & ($10^{51}$\,erg) & (\msun)              & (\msun) & (\msun) & (\msun) & (\msun) & (\msun)  & (\msun) \\
\hline
toy06 & 1.00 & 1.00 & 0.600 & 0.001 & 0.040 & 0.140 & 0.220 & 0.000 & 0.000 \\
toy01 & 1.00 & 1.00 & 0.100 & 0.000 & 0.090 & 0.315 & 0.495 & 0.000 & 0.000 \\
DDC10 & 1.42 & 1.51 & 0.620 & 0.112 & 0.041 & 0.166 & 0.257 & 0.101 & 0.002 \\
DDC25 & 1.39 & 1.18 & 0.115 & 0.098 & 0.024 & 0.237 & 0.478 & 0.282 & 0.022 \\
\hline
\end{tabular}
\end{center}
\end{table*}

Comparisons of the results of different RT codes for the same ejecta
models play an important role in estimating the accuracy of different
approximations and can be used to validate new codes.  The number and
sophistication of RT codes has significantly developed in recent
years, increasing the need for diverse benchmark ejecta models that
will allow detailed and careful comparisons. In this paper, we describe
the first results of a collaborative effort of ten groups around the
world that are developing existing RT codes (in alphabetical order:
ARTIS, CMFGEN, CRAB, KEPLER, SEDONA, STELLA, SUMO, SuperNu, TARDIS,
and URILIGHT; see Sect.~\ref{sect:codes} for descriptions and
references) to create a systematic code-comparison framework. As a
first important step, all groups agreed on a set of four test model
ejecta and standardised output formats. Each group calculated the
resulting radiative display with their respective codes for the same
ejecta models and shared the results in a new public electronic
repository on
GitHub\footnote{\url{https://github.com/sn-rad-trans}}. We did not
attempt to agree on a specific setup for each code, nor did we
synchronise atomic data between codes.

The structure of the paper is as follows. In Sect.~\ref{sect:models},
the benchmark models are described. These include two simplistic
sub-Chandrasekhar-mass (sub-\mch) toy models with profiles that are
defined analytically and two more realistic models that result from
hydrodynamical simulations of the \mch\ delayed-detonation
scenario. All the models considered here are spherically symmetric
(1D). We give short descriptions of the RT codes that were employed in
this first comparison in Sect.~\ref{sect:codes}, with emphasis on the
main physical approximations that are used in each. In
Sect.~\ref{sect:outputs} we describe the publicly available repository
of results. In particular, descriptions are provided of the output
files and of Python codes that are included for reading them. In
Sect.~\ref{sect:results}, several example comparisons of the results
of the different codes are provided in order to illustrate the
available outputs. We voluntarily make no attempt to analyse the
sources of discrepancies. While the comparisons focus on observable
aspects of the emission, comparisons to observations and conclusions
regarding the implications for the applicability of the codes are
intentionally not addressed in order to keep the focus of the paper on
the description of the comparison. Finally, we outline future plans
for this comparison project in Sect.~\ref{sect:conclusions}.

%%%%%%%%%%%%%%%%%%%%%%%%%%%%%%%%%%%%%%%
% Test Models
%%%%%%%%%%%%%%%%%%%%%%%%%%%%%%%%%%%%%%%

\section{Test models}\label{sect:models}

\begin{figure}
 \includegraphics[width=\hsize]{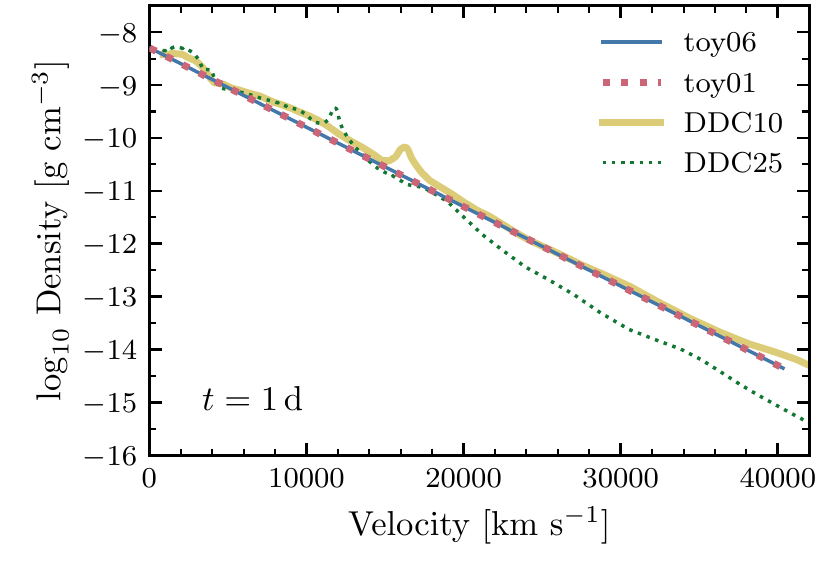}
 \caption{Density profiles of our model set at an adopted time of 1\,d post explosion.
 }
 \label{fig:dens}
\end{figure}

\begin{figure*}
 \includegraphics[width=0.5\hsize]{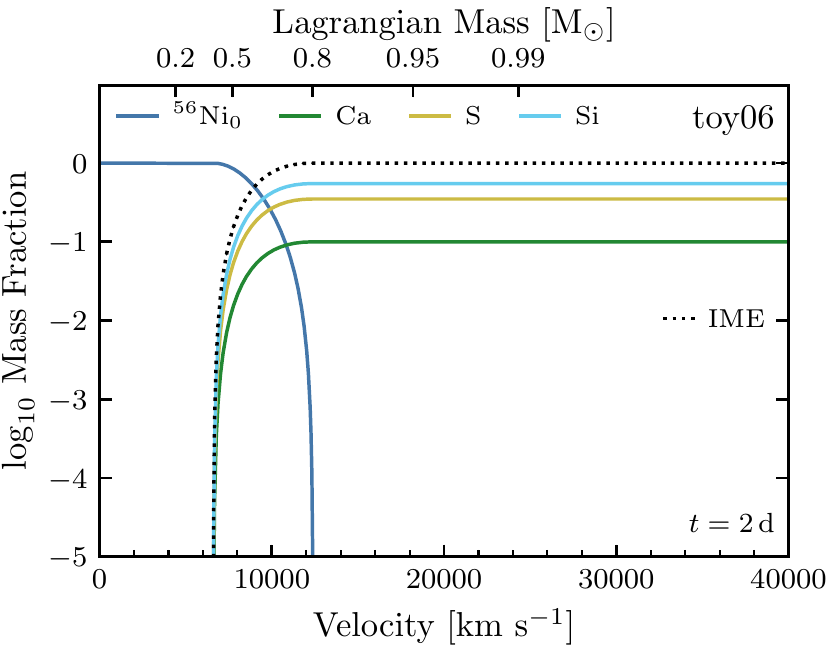}
 \includegraphics[width=0.5\hsize]{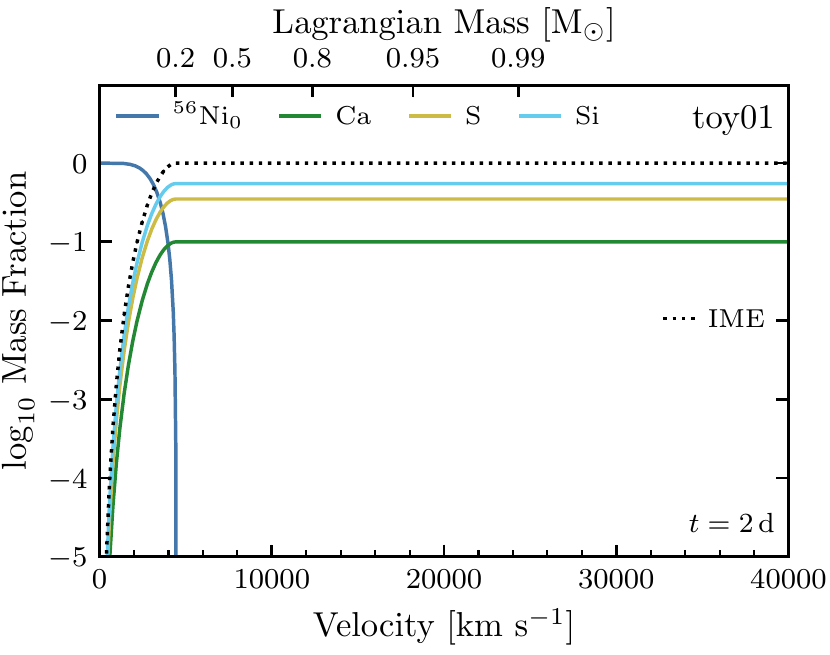}
 \includegraphics[width=0.5\hsize]{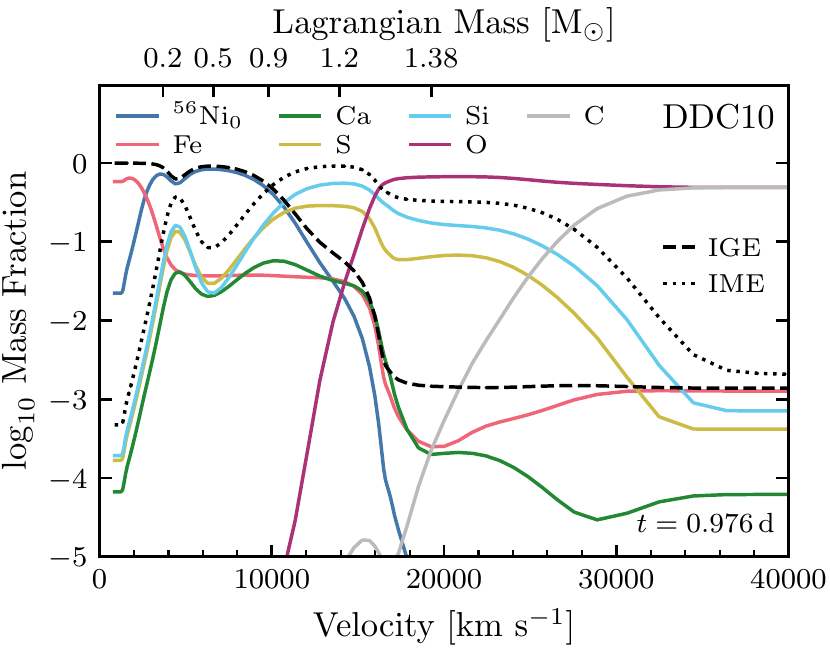}
 \includegraphics[width=0.5\hsize]{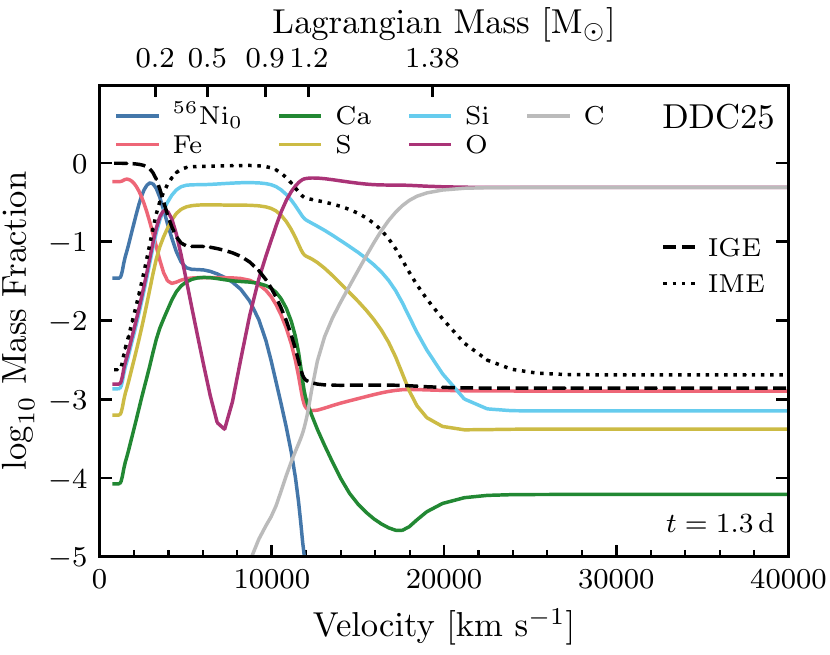}
 \caption{Composition profiles at the start of our simulations in our
   model set (2\,d post explosion for the toy models and $\sim$1\,d
   post explosion for the DDC models). For the toy models, this
   represents the full set of species present (\nifs\ and decay
   products \cofs\ and \fefs, as well as IMEs: Ca S, Si), while for
   the DDC models only a subset of species are shown for
   illustration. The \nifs\ mass fraction is given at the time of
   explosion. We also show the total IGE mass fraction (from Sc to Ni;
   dashed line) and the total IME mass fraction (from Na to Ca; dotted
   line). The total IGE mass fraction coincides with the $\nifs_0$
   line in the toy models and is not shown for sake of clarity.
 }
 \label{fig:composition}
\end{figure*}

The code-comparison test suite consists of four \snia\ models. Two are
sub-\mch\ models with analytic density and composition profiles
(Sect.~\ref{sect:toymodels}; `toy' models), and the remaining two
are \mch\ models resulting from hydrodynamical simulations
(Sect.~\ref{sect:ddcmodels}; DDC models). The models were set up or
selected based on their \nifs\ yield, in order to have two models
corresponding to `normal' \sneia\ (toy06 and DDC10 with
$\sim$0.6\,\msun\ of \nifs) and two low-luminosity models (toy01 and
DDC25 with $\sim$0.1\,\msun\ of \nifs). We present the toy and DDC
models in turn in the following sections, and summarise their
properties in Table~\ref{tab:models}. The density profiles at a
reference time of 1\,d post explosion and initial composition profiles
are shown in Figs.~\ref{fig:dens} and \ref{fig:composition},
respectively.

\subsection{Toy models}\label{sect:toymodels}

The toy models were generated using the \verb+mk_snia_toy_model.py+
script (see Sect.~\ref{sect:outputs}) using the \verb+--highni+ (for
the normal \snia\ model) and \verb+--lowni+ (for the low-luminosity
\snia\ model) options. Both models have a total mass
$M_\mathrm{ej}=1.0$\,\msun, a kinetic energy
$E_\mathrm{kin}=10^{51}$\,erg, and are calculated at the time
$t_f=2$\,d post explosion. The models have an exponential density
profile \citep[e.g.][]{Jefferey1999},
\begin{equation}\label{eq:rhoexp}
  \rho = \rho_\mathrm{c}e^{-v/v_e},
\end{equation}
where
\begin{equation}\label{eq:ve}
  v_e = \sqrt{\frac{E_\mathrm{kin}}{6M_\mathrm{ej}}} \approx 2895\mathrm{\,\kms}
\end{equation}
is the e-folding velocity, and 
\begin{equation}\label{eq:rhoc0}
  \rho_\mathrm{c} = \frac{M_\mathrm{ej}}{8\pi v_e^3 t_f^3} \approx 6.32 \times 10^{-10}\mathrm{\,\gcc}
\end{equation}
is the central density at the chosen time.

The ejecta consist of $n$ spherical shells indexed by
$i=1,\ldots,n$. We use a uniform velocity grid with width $\Delta v =
50$\,\kms. Each shell then has an inner and outer velocity coordinate
given by: $v_{0,i} = (i-1) \Delta v$ and $v_{1,i} = v_{0,i} + \Delta
v$.

As in \cite{Jefferey1999}, we define the dimensionless radial
coordinate $z = v / v_e$ which we use to compute the mass of each
shell as:

\begin{equation}\label{eq:mshell}
  M_i = \int \rho(z)\ \mathrm{d}V = \rho_\mathrm{c} \int e^{-z}\ \mathrm{d}V,
\end{equation}

\noindent
where the integral runs from $z_{0,i}$ to $z_{1,i}$ and the volume
element is $dV = 4\pi r^2 dr = 4\pi v_e^3 t_f^3 z^2 dz$, where we
assume a homologously expanding ejecta ($r = vt = v_e z t$).

Likewise, each shell volume $V_i$ is computed from the inner and outer
radii ($r_{i,\{0,1\}} = v_{i,\{0,1\}} t_f$), which results in the mean
density of each shell:

\begin{equation}\label{eq:rhoave}
  \rho_i = \frac{M_i}{V_i} = 3 \rho_c \frac{
    e^{-z_{0,i}}(z_{0,i}^2 + 2z_{0,i} + 2) -
    e^{-z_{1,i}}(z_{1,i}^2 + 2z_{1,i} + 2)}
      {z_{1,i}^3 - z_{0,i}^3}.
\end{equation}

\noindent
Starting from the central shell, we keep adding successive shells
until the total mass is within 0.01\% of the required ejecta mass.
For the 1.0\,\msun\ toy models considered here, this results in 807
shells, where the last shell has an outer velocity of 40\,350\,\kms.
We note that not all codes use this velocity grid; in particular, the
number of shells and maximum velocity vary significantly among codes
(see Sect.~\ref{sect:codes}).

We assume the ejecta consist of a central region composed of
iron-group elements (IGEs), here only
\nifs\ and its decay products \cofs\ and \fefs, and an outer region
composed of the intermediate-mass elements (IMEs) Ca, S, and Si with
constant mass fractions throughout the layer: $X(\mathrm{Ca})=0.1$,
$X(\mathrm{S})=0.35$, and $X(\mathrm{Si})=0.55$. These values were
chosen based on the delayed-detonation model DDC10 (see
Sect.~\ref{sect:ddcmodels} below). Our `normal' toy model consists of
0.6\,\msun\ of \nifs\ and 0.4\,\msun\ of IMEs, while our
low-luminosity toy model consists of 0.1\,\msun\ of \nifs\ and
0.9\,\msun\ of IMEs.  The \nifs\ and IME composition profiles are
smoothly connected using an analytic function (here a cosine bell)
over a mass interval $\Delta M_\mathrm{trans}$ (set to 0.2\,\msun\ for
the low-luminosity model and 0.4\,\msun\ for the regular model). At a
given mass coordinate $m$, the \nifs\ mass fraction is set to:

\begin{equation}
  X_{\nifs}(m) = 
  \begin{cases}
    1                     & \text{for $m < m_1$}, \\
    1 - f_\mathrm{cos}(m) & \text{for $m_1 \le m \le m_2$}, \\
    0                     & \text{for $m > m_2$},
  \end{cases}
\end{equation}

\noindent
where $m_1 = M(\nifs) - \Delta $$M_\mathrm{trans}/2$, 
$m_2 = M(\nifs) + \Delta $$M_\mathrm{trans}/2$, and

\begin{equation}
f_\mathrm{cos}(m) = \dfrac{1}{2} \left\{
1 - \cos \left[ \left( \dfrac{m - m_2}{\Delta M_\mathrm{trans}} \right) \pi \right]
\right\}.
\end{equation}

\noindent
The IME mass fraction is then simply set to $X_\mathrm{IME}(m) = 1 -
X_{\nifs}(m)$.  Our toy models therefore consist of only six chemical
species or isotopes (\nifs, \cofs, \fefs, $^{40}$Ca, $^{32}$S,
$^{28}$Si): this was a voluntary choice in order to keep the model as
simple as possible while retaining the defining characteristics of a
\snia.

The initial temperature profile at $t_f=2$\,d is determined by solving
the first law of thermodynamics assuming a radiation-dominated gas,
local energy deposition from \nifs\ decay, and no diffusion (i.e. the
temperature in each zone is solved independently of the adjacent
zones). Given these assumptions, the temperature at $t_f$ can be
determined analytically by noting that the time-weighted internal
energy, $tE(t)$, is equal to the time-integrated time-weighted decay energy
deposition rate, $\int{t Q(t) \mathrm{d}t}$, as noted by
\cite{Katz2013}:

\begin{equation}\label{eq:temp}
  T_i = \left( \frac{\rho_i X_i(\mathrm{\nifs})_0 \int_0^{t_f} tq_{\rm Ni}(t) \mathrm{d}t}{a ~ t_f} \right) ^{1/4},
\end{equation}

\noindent
where $X_i(\nifs)_0$ is the \nifs\ mass fraction at $t\approx0$ in the
i'th cell, $a$ is the radiation constant, and $q_{\rm Ni}(t)$ is the
energy release rate per unit mass (ignoring neutrinos) of the
\nifs$\rightarrow$\cofs$\rightarrow$\fefs\ decay chain.  In this
formulation, we ignore the time-weighted internal energy shortly after
explosion, $E(t_0) t_0$.

It is clear from Eq.~\ref{eq:temp} that the temperature is predicted
to be zero in zones devoid of \nifs\ ($\gtrsim 12000$\,\kms). We therefore
impose a constant temperature floor in these zones of 5000\,K.  The
resulting excess internal energy is quickly radiated away because these
layers have a relatively low optical depth, such that the impact on
the longer-term radiative display is negligible.

\subsection{Delayed-detonation models}\label{sect:ddcmodels}

In addition to the two toy models above, we consider two
\mch\ delayed-detonation models resulting from hydrodynamical
simulations. These were chosen from the DDC model grid presented in
\cite{Blondin2013} in order to closely match the \nifs\ yields of the
toy models: Our low-luminosity model DDC25 yields $\sim
0.12$\,\msun\ of \nifs\ (cf. 0.1\,\msun\ for the toy01 model), and our
`normal' DDC10 model yields $\sim 0.52$\,\msun\ of
\nifs\ (cf. 0.6\,\msun\ for the toy06 model). We refer the reader to
\cite{Blondin2013} for a detailed description of the DDC models.

The outputs of the hydrodynamical modelling correspond to 30-60\,s post explosion, by which time the ejecta have reached a state of homologous
expansion. We applied a small amount of radial mixing to the hydrodynamical
input with a characteristic velocity width $\Delta v_{\rm
  mix}=400$\,\kms\ to smooth sharp variations in composition.  The
ejecta were then evolved to 0.5\,d post explosion by solving the
energy equation given by the first law of thermodynamics, assuming the
plasma is radiation dominated and neglecting diffusion. Apart from the
two-step \nifs$\rightarrow$\cofs$\rightarrow$\fefs\ decay chain, we
also treat eight additional two-step decay chains associated with
$^{37}$K, $^{44}$Ti, $^{48}$Cr, $^{49}$Cr, $^{51}$Mn, $^{52}$Fe,
$^{55}$Co, and $^{57}$Ni, and a further six one-step decay chains
associated with $^{41}$Ar, $^{42}$K, $^{43}$K, $^{43}$Sc, $^{47}$Sc, and
$^{61}$Co \citep[see][]{Dessart2014}.

The ejecta at 0.5\,d are then remapped onto the 1D, non-LTE,
radiative-transfer code CMFGEN of \cite{Hillier2012_cmfgen} and
evolved until $\sim 1$\,d post explosion (0.976\,d for the DDC10 model
and 1.3\,d for the DDC25 model; see Sect.~\ref{sect:cmfgen} for
details), at which point the ejecta serve as initial conditions for
the other radiative-transfer codes presented in
Sect.~\ref{sect:codes}.

%%%%%%%%%%%%%%%%%%%%%%%%%%%%%%%%%%%%%%%
% Codes
%%%%%%%%%%%%%%%%%%%%%%%%%%%%%%%%%%%%%%%

\section{Radiative-transfer codes}\label{sect:codes}

%%% Table: code physical ingredients
\begin{table*}
\centering
\scriptsize
\caption{Physical ingredients and approximations used in each code for the test models in this paper.}\label{tab:codephys}
\begin{tabular}{lccccccccc}
\hline
Code          & RT method     & Homologous   & $\gamma$-ray   & Non-thermal      & Excitation         & Ionisation         & Radiation           & Line Opacity               & Thermalisation         \\
              &               & expansion    & deposition     & deposition       &                    &                    & field $J_\nu$       & $\kappa_\nu$               & parameter $\epsilon$   \\
(1)           & (2)           & (3)          & (4)            & (5)              & (6)                & (7)                & (8)                 & (9)                        & (10)                   \\
\hline                                                                                                                                                                                                
ARTIS         & MC            & yes          & MC             & $\cdots$         & LTE($T_R$)         & approx. $dn/dt=0$  & scaled LTE($T_R$)   & Sobolev                    & $\cdots$               \\ %Macro Atom (bf,ff) + Sobolev (bb)
ARTIS nebular & MC            & yes          & MC             & Spencer-Fano     & $dn/dt=0$          & $dn/dt=0$          & $dJ/dt$             & Sobolev                    & $\cdots$               \\ 
CMFGEN        & RTE-CMF       & yes          & MC             & Spencer-Fano     & $dn/dt$            & $dn/dt$            & $dJ/dt$             & $\kappa_\nu$               & $\cdots$               \\
CRAB          & RH-1G         & no           & grey           & Kozma/Fransson   & LTE($T_R$)         & LTE($T_R$)         & $dJ/dt$             & Expansion                  & 0.9                    \\ 
KEPLER        & FLD           & no           & grey           & $\cdots$         & $\cdots$           & LTE($T_e$)         & LTE($T_e$)          & $\kappa=\mathrm{const.}$   & $\cdots$               \\ %Kappa_add=0.1 cm^2/g
SEDONA        & MC            & yes          & MC             & $\cdots$         & LTE($T_e$)         & LTE($T_e$)         & $dJ/dt$             & Expansion                  & 0.8 or 1.0   \\
SUMO          & MC            & yes          & grey           & Spencer-Fano     & $dn/dt=0$          & $dn/dt=0$          & $dJ/dt=0$           & Sobolev                    & $\cdots$               \\
STELLA        & RH-MG         & no           & grey           & $\cdots$         & LTE($T_e$)         & LTE($T_e$)         & $dJ/dt$             & Expansion                  & 0.9                    \\
SuperNu       & MC            & yes          & MC grey        & $\cdots$         & LTE($T_e$)         & LTE($T_e$)         & $dJ/dt$             & $\kappa_\nu$               & 1.0                    \\
TARDIS        & MC            & yes          & $\cdots$       & $\cdots$         & scaled LTE($T_R$)  & scaled LTE($T_R$)  & scaled LTE($T_R$)   & Sobolev                    & $\cdots$               \\
URILIGHT      & MC            & yes          & MC             & $\cdots$         & LTE($T_e$)         & LTE($T_e$)         & $dJ/dt$             & Expansion                  & 0.8                    \\
\hline
\end{tabular}
\flushleft
\textbf{Notes:} 
\textit{Column headings:}
(1) Code name.
(2) Numerical method used to solve the
radiative-transfer equation: FLD=Flux Limited Diffusion, MC=Monte
Carlo, RH-1G = one-group (grey) radiation hydrodynamics, RH-MG =
multi-group radiation hydrodynamics, RTE-CMF= Radiation Transfer
Equation Co-Moving Frame.
(3) The ejecta are assumed to be in homologous expansion ($v=rt$) in radiative-transfer codes. This is not the case for radiation-hydrodynamics codes (CRAB, KEPLER, STELLA).
(4) Treatment of $\gamma$-ray energy deposition.
(5) Non-thermal heating, excitation, and ionisation rates are calculated through a solution of the Spencer-Fano equation \citep{Spencer1954} or read in from tabulated values \citep{Kozma1992}.
(6) Solution method for the atomic level populations. LTE($T_X$) refers to a solution of the Boltzmann excitation formula setting the temperature to that of the electrons ($T_e$) or the radiation field ($T_R$). An approximate non-LTE treatment of excitation scales the Boltzmann occupation numbers by the dilution factor $W$ (cf. dilute-LTE treatment in TARDIS; Sect.~\ref{sect:tardis}). A non-LTE treatment requires the solution of the rate equations, either including time dependence ($dn/dt$) or assuming steady-state (statistical equilibrium, $dn/dt=0$).
(7) Treatment of ionisation. Here LTE($T_X$) refers to a solution of the Saha-Boltzmann equation, which can be scaled for an approximate non-LTE treatment (cf. nebular approximation in TARDIS; Sect.~\ref{sect:tardis}). The non-LTE solution results from the solution of the rate equations, either including time dependence ($dn/dt$) or assuming steady-state ($dn/dt=0$).
(8) The radiation field can be computed via a solution of the radiative-transfer equation (possibly assuming steady-state, $dJ/dt=0$) or by following the propagation of photon packets in Monte Carlo codes. Alternatively, LTE treatments assume a Planckian radiation field (black body $B_\nu$) at a reference temperature $T_X$, possibly scaled by the dilution factor $W$.
(9) Treatment of line opacity. This can be explicitly line by line, taking into account overlap in the co-moving frame ($\kappa_\nu$), or with use of the Sobolev approximation. Other treatments involve the use of an approximate frequency-dependent `expansion' opacity, or assuming a constant value (e.g. KEPLER; Sect.~\ref{sect:kepler}).
(10) Global value of the thermalisation parameter $\epsilon$, which sets the probability that a photon absorbed in a given transition is re-emitted in a different transition (see e.g. URILIGHT; Sect.~\ref{sect:urilight}).
\end{table*}

In the following subsections, each group provides a brief description
of their code, highlighting the specific setup used in the
calculations for this paper. Each code subsection follows a similar
structure: brief description of the code (and whether it assumes a
homologous velocity law); treatment of $\gamma$-ray energy deposition
and non-thermal effects; computation of the temperature structure;
treatment of excitation and ionisation; evaluation of the radiation
field; sources of opacity and atomic data; resolution (spatial and
frequency) and typical runtime. Table~\ref{tab:codephys} summarises
the physical ingredients and approximations used in each code for the
test models considered here.

\subsection{ARTIS}\label{sect:artis}

ARTIS\footnote{Source code available at
\url{https://github.com/artis-mcrt/artis}} is a Monte Carlo radiative
transfer code that uses the indivisible energy packet method of
\cite{Lucy2002}. The code was originally described by \cite{Sim2007}
and \cite{Kromer2009}, with later improvements presented by
\cite{Bulla2015} and \cite{Shingles2020}.  The code is
three-dimensional and follows the time-evolution of the radiation
field and state of the gas.  It assumes a strictly homologous velocity
law.

Injection of energy into the ejecta is calculated by following the
deposition of $\gamma$-ray packets that are injected in accordance
with the radioactive decays of $^{56}$Ni and $^{56}$Co, following
\cite{Lucy2005}. Additional decay channels have been included in the
studies of specific models. The simulated $\gamma$-ray transport is
non-grey and takes into account Compton scattering, photoelectric
absorption, and pair-creation opacities. In our standard runs, the code
does not include the effects of excitation or ionisation by
non-thermal particles. However, \cite{Shingles2020} presented updates
to the code that include a Spencer-Fano treatment of non-thermal
ionisation as required for late-phase modelling. Results obtained
with this improved version (\texttt{artisnebular}) are included for
late phases for the models in this study.

The electron temperature in each grid zone is estimated by balancing
of heating and cooling rates (accounting for $\gamma$-ray and positron
deposition, bound-bound, bound-free, and free-free processes).  In its
regular mode of operation (\texttt{artis}), the code uses an
approximate non-LTE treatment to estimate the ionisation state in the
ejecta (based on Monte Carlo photoionisation estimators; see
\citealt{Kromer2009}) and an LTE treatment of excitation is
adopted. This approach has been used in most of our published studies,
and is also used in most of the \texttt{artis} calculations presented
here. However, this method has limitations that become increasingly
important at later phases (e.g. it neglects non-thermal
heating and ionisation and tends to overestimate the plasma temperature at
low densities due to incomplete treatment of cooling by forbidden
lines). To improve these issues, \cite{Shingles2020} presented updates
to the code that include a full non-LTE population solver (together
with the Spencer-Fano solver mentioned above). Results obtained with
this improved version (\texttt{artisnebular}) are included for
late-phase calculations here.

Monte Carlo estimators are used to track the radiation field in each
grid cell. In general, we use volume-based estimators (see
\citealt{Lucy1999} or \citealt{Noebauer2019}) to extract radiation-field-dependent quantities from the flight histories of our Monte
Carlo quanta. In its standard mode of operation, \texttt{artis} uses
detailed Monte Carlo estimators to obtain photoionisation rates from
the radiation field, but relies on an estimated scaling for excited-state photoionisation and on a dilute black-body radiation field model
for bound-bound excitation (see \citealt{Kromer2009} for details). However, the
improved \texttt{artisnebular} version uses a more detailed
frequency binned representation of the radiation field (see
\cite{Shingles2020} for details). The code has the capacity to iterate
on each time step with the aim of achieving consistency between the
radiation field estimates and the packet transport in each step. However, in
practise we find that this iteration is not required and we
therefore generally simply use the radiation field quantities
extracted from the previous time step to estimate the radiative rates
that are needed for the current step.

In simulating ultraviolet to infrared photon transport, the code
accounts for electron scattering, bound-bound, bound-free, and
free-free processes. Bound-free and bound-bound processes are treated
using the \emph{Macro Atom} approach of \cite{Lucy2002, Lucy2003} and
adopting the Sobolev approximation for line opacity. The code does not
use an expansion opacity (or similar) but treats line opacity based on
a frequency-ordered list of transitions treated in the Sobolev limit
(i.e. no line overlap is taken into account).

In our simulations, atomic data are primarily drawn from the Kurucz
atomic line lists (see \citealt{Kromer2009}) -- in
Appendix~\ref{sect:atomdata_artis} we summarise the ions and numbers
of levels and lines that we include.  The photon transport is carried
out on a 3D Cartesian grid that co-expands with the ejecta. The
\texttt{artis} simulations included here were carried out on a 100$^3$
grid. The resolution therefore corresponds to around 500 -- 1000 km
s$^{-1}$, depending on the model. The simulations are typically run on
1000 computer cores for one to two days.

\subsection{CMFGEN}\label{sect:cmfgen}

CMFGEN is a 1D, non-LTE, time-dependent radiative-transfer code that
solves the transfer equation in the co-moving frame of spherical
outflows.  Details about the code, techniques, and approximations can
be found in \cite{Hillier1998_cmfgen},
  \cite{Hillier2003_cmfgen}, \cite{Hillier2012_cmfgen_HSW},
and (for SN calculations) in
\cite{Hillier2012_cmfgen}\footnote{CMFGEN, with documentation, is
available at \url{www.pitt.edu/~hillier}}.  The velocity law for the
outflow is in general monotonic (but see e.g. \citealt{Dessart2015}
in the case of interacting SNe) and is assumed here to be
homologous (such that $\partial v/\partial r = v/r$).
 
In the present calculations, non-local energy deposition from
radioactive decay is treated using a Monte-Carlo approach for
$\gamma$-ray transport \citep{Hillier2012_cmfgen}. Non-thermal
processes associated with the high-energy electrons produced by
Compton scattering and photoelectric absorption of these $\gamma$ rays
are accounted for through a solution of the Spencer-Fano equation
\citep{Spencer1954,Li2012}.

The temperature structure is constrained through the energy equation
that has the form:
\begin{equation}
\rho {De \over Dt} - {P \over \rho} {D\rho \over Dt}= 4\pi \int d\nu
(\chi_\nu J_\nu - \eta_\nu) + \dot \epsilon_{\rm decay} \,,
\label{eq_energy}
\end{equation}
where $D \over Dt$ is the Lagrangian derivative, \noindent $e$ is the
internal energy per unit mass, $P$ is the gas pressure, and $\dot
\epsilon_{\rm decay}$ is the energy absorbed per second per unit
volume (see \citealt{Hillier2012_cmfgen} for further details).  We
verify the accuracy of the solution by examining a global energy
constraint (equivalent to conservation of flux in a static atmosphere;
see \citealt{Hillier2012_cmfgen} for details), and an equation
describing energy conservation as applied to the heating and cooling
of free electrons. These two equations are related to the above
Eq.~\ref{eq_energy} via the transfer equation, and the rate equations,
respectively \citep[e.g.][]{Hillier2003_cmfgen}. Because of model
assumptions (e.g. the use of super-levels) these two equations are not
satisfied exactly, but the errors (typical at the 1\% level or
smaller) are too small to affect the validity of the models.
Processes contributing to electron heating and cooling include
bound-free ionisation and recombination, collisional ionisation and
recombination, collisional excitation and de-excitation, free-free
emission, Auger ionisation, charge exchange reactions (primarily with
H\one\ and He\one, and hence negligible in \snia\ ejecta), net cooling
from non-thermal processes, and heating by radioactive decay.

Atom and ion-level populations are determined through a solution of
the time-dependent rate equations, coupled to the above energy
equation and the zeroth and first moments of the
radiative-transfer equation (see below). We consider the following
processes: bound–bound processes, bound–free processes, collisional
ionisation and recombination, collisional excitation and
de-excitation, Auger ionisation
\citep{Hillier1987,Hillier1998_cmfgen}, and non-thermal excitation and
ionisation \citep{Li2012}.  We additionally consider further processes
involving H and He (two-photon decay, charge-exchange reactions, and
Rayleigh scattering), although these are negligible here (and
completely absent from the toy models, which contain no H or He).  To
ease the solution of the rate equations, atomic levels are grouped
into super levels (see \citealt{Hillier1998_cmfgen} for details).

The frequency-dependent mean intensity $J_\nu$ is obtained via a
solution of the time-dependent transfer equation in the co-moving
frame to first order in $v/c$. More specifically, we solve the zeroth and first moment equations, which are closed using
so-called Eddington factors $f_\nu=K_\nu/J_\nu$, where $K_\nu$ is the
second moment of the specific intensity (related to the radiation
pressure). The Eddington factors are obtained from a formal solution
of the time-independent transfer equation.

We consider the following sources of opacity: electron scattering,
bound-free (including photoionisation from excited states),
bound-bound\footnote{In these calculations we assume a Doppler profile
with a constant effective Doppler width (including both thermal and
turbulent velocities) of 50\,\kms.  In more general SN
modelling, the effective Doppler width is varied to test its effect on
model results.}, free-free, and Auger ionisation. As noted earlier,
Rayleigh scattering and two-photon processes (for H and He only) are
also part of the global opacity budget but are negligible here.

A description of the sources of atomic data can be found in the
Appendix (Sect.~\ref{sect:atomdata_cmfgen}).  The number of levels
(both super-levels and full levels) and corresponding number of
bound-bound transitions are given in
Tables~\ref{tab:cmfgen_atoms_toy06}-\ref{tab:cmfgen_atoms_ddc_reducednicofe4}.
We ignore weak transitions with a $gf$ value\footnote{This value is
the product of the statistical weight $g$ of the lower level and the
oscillator strength $f$ of an atomic transition.} below some cutoff,
typically set to $10^{-4}$.  For the toy models, the following ions
were included: Si\two--{\sc iv}, S\two--{\sc iv}, Ca\two--{\sc iv},
Fe\one--{\sc v}, Co\two--{\sc vii}, and Ni\two--{\sc v}.  For the
delayed-detonation DDC models the following ions were included:
C\one--{\sc iv}, O\one--{\sc iv}, Ne\one--{\sc iii}, Na\one,
Mg\two--{\sc iii}, Al\two--{\sc iii}, Si\two--{\sc iv}, S\two--{\sc
  iv}, Ar\one--{\sc iii}, Ca\two--{\sc iv}, Sc\two--{\sc iii},
Ti\two--{\sc iii}, Cr\two--{\sc iv}, Mn\two--{\sc iii}, Fe\one--{\sc
  vii}, Co\two--{\sc vii}, and Ni\two--{\sc vii} (we also include the
ground states of Cl\four, K\three, and V\one\ for the sole purpose of
tracking changes in the abundances of radioactive isotopes). For all the
aforementioned ions, we also consider ionisations to and
recombinations from the ground state of the next ionisation stage
(e.g. \ion{Ni}{8} in the case of Ni). As time proceeds and the
temperature in the spectrum-formation region drops, the highest
ionisation stages have a negligible impact on the RT solution.  When
this occurs, smaller model atoms are used for these ions, or the ions
are removed altogether from the atomic model set.

The toy models were remapped onto a coarser spatial grid with 100
depth points. No remapping was performed for the DDC models.  Typical
wall-clock runtimes are of the order of 24h per time step on a single
computing node with 8-12 CPUs, thus taking 2-3 months to complete for
a sequence covering the first 200\,d or so post explosion.

\subsection{CRAB}\label{sect:crab}

CRAB is a 1D, implicit, Lagrangian radiation hydrodynamics code developed
to study the light curves during SN outbursts and the corresponding
outflows evolved from hydrostatic state up to homologous expansion
\citep{Utrobin2004}.
Non-local energy deposition of $\gamma$ rays from radioactive decay is
determined by solving the corresponding one-group $\gamma$-ray transport
with the approximation of an effective absorption opacity of 0.06
$Y_\mathrm{e}$\,cm$^2$\,g$^{-1}$. Positrons are assumed to deposit
their energy locally. This energy deposition produces non-thermal
ionisation and heating, the rates of which are taken from \cite{Kozma1992}.

The radiation hydrodynamic equations include a time-dependent energy
equation, which is based on the first law of thermodynamics and determines
the gas temperature structure. In the outer, transparent and semitransparent
layers of the SN ejecta, the local energy balance is in control of a net
balance between heating and cooling processes, while in the inner, optically thick
layers, it is determined by the diffusion of equilibrium radiation.

The code has two options for the treatment of atom and ion level
populations. In option A, the elements H, He, C, N, O, Ne, Na, Mg, Si, S,
Ar, Ca, and Fe, and the negative hydrogen ion H$^{-}$ are included in the
non-LTE ionisation balance.  All elements but H are treated with three
ionisation stages.  The ionisation balance is controlled by the
following elementary processes: photoionisation and radiative
recombination, electron ionisation and three-particle recombination,
and non-thermal ionisation. In option B, we use an LTE treatment of ionisation
and excitation for elements from the option A list or all elements
from H to Zn modelled with an arbitrary number of ionisation stages. Atomic and ionic level populations are determined using the
Boltzmann formulae and the Saha equations for an element mixture
with the local electron or non-equilibrium radiation temperature.

The time-dependent radiation transfer is treated in a one-group (grey)
approximation in the outer, transparent and semitransparent layers of
the SN envelope, while in the inner, optically thick layers where
thermalisation of radiation takes place and LTE applies, the diffusion
of equilibrium radiation is described in the approximation of radiative
heat conduction. The bolometric luminosity of the SN is calculated by
including retardation and limb-darkening effects.

In the inner, optically thick layers, the Rosseland mean opacity is
evaluated for the local electron temperature, while in the outer,
transparent and semitransparent layers non-LTE effects are taken into
account when determining the mean opacities and the thermal emission
coefficient. The mean opacities include processes of photoionisation,
free-free absorption, Thomson scattering on free electrons, and
Rayleigh scattering on neutral hydrogen. The contribution of lines to
the expansion opacity is evaluated by the generalised formula of
\citet{Castor1975} or by the formula of \citet{Blinnikov1996_opac}
using the Sobolev approximation for line opacity.  The expansion opacity
in an expanding medium is treated with a thermalisation parameter set
to 0.9 as recommended by \cite{Kozyreva2020} to model \sneia.

The sources of atomic data for processes in continuum can be found in
\citep{Utrobin2004}. Oscillator strengths of lines are taken from the Kurucz
line database\footnote{\url{http://kurucz.harvard.edu/cdroms.html}}
containing nearly 530\,000 lines. Energy level data are from the
atomic spectra database of the National Institute of Standards and
Technology.

The zoning of model toy06 with 808 zones is adequate for modelling a light curve.
For this model, a typical runtime is of the order of 5 h on one CPU for
the first 140\,d.

\subsection{KEPLER}\label{sect:kepler}

KEPLER is a one-dimensional, implicit Lagrangian hydrodynamics code
with appropriate physics for the study of massive stellar evolution and
SNe \citep{Weaver1978,Woosley2002}. The radiation transport is
flux-limited diffusion using a single temperature for the matter and
radiation.

An important difference between KEPLER and some of the other codes
used here is that KEPLER does not assume a coasting configuration.
At early times in particular, when the matter is very optically thick,
energy input from $^{56}$Ni and $^{56}$Co decay will both heat the
matter and accelerate it. The correction to the kinetic energy is
small for the cases studied here, but the integral of the emitted
light will be less than the integral of the decay energy that is
deposited (minus adiabatic losses).

Our approach to $\gamma$-ray energy deposition is discussed in the
Appendix of \citet{Ensman1988}.  Since the early 1990s, a value of
$\kappa_{\gamma}$ = 0.054 cm$^{-2}$ g$^{-1}$ has been used for the
global opacity parameter in KEPLER's leakage scheme to model
\sneia\ and that is the value used here.  A better value can be
obtained by comparing with Monte Carlo calculations for a given class of
model, such as SN IIP, SN Ib, and so on.  The value 0.054 cm$^2$ g$^{-1}$ is
larger than the standard local grey opacity $(0.06 \pm 0.01) Y_e$
cm$^2$ g$^{-1}$ \citep[e.g.][]{Swartz1995}, where $Y_e \approx 0.5$,
as it is used to calculate the effective column depth from the
outer edge of a spherical zone to the surface. For all points except
the geometric centre of the explosion, the angle-averaged column depth
would be greater than along this radial line. The averaging is therefore
approximated by taking a larger opacity. In reality, this number would
vary with the radial distribution of $^{56}$Ni and would be smaller if
all the $^{56}$Ni were at the centre.

The temperature structure is computed by solving the hydrodynamics
equations including the effects of expansion and acceleration with
energy input by radioactive decay and transported by radiative
diffusion. Because of the way KEPLER handles flux-limited transport using a
single temperature for the radiation and matter, the pressure in the
outermost zones can be overestimated. This can lead to a small (of
order 10\%) overestimate of the conversion of radiation to kinetic
energy in those zones at late times. To alleviate this problem, the
luminosity is taken to be the maximum of the value at the
(electron-scattering) photosphere and the zone that includes 95\% of
the mass. The former dominated the light curve until after peak.

An important parameter of the calculation is the {atomic} opacity used to transport thermal radiation. In KEPLER, the
total `optical' opacity consists of two parts: (a) electron scattering,
which is calculated using a full Saha solve for all ionisation stages
of all 19 species present; and (b) a constant additive opacity,
$\kappa_a$, taken to reflect the effect of Doppler-broadened
lines. The electron-scattering opacity is temperature-, density-, and
composition dependent and therefore varies with location and time. The
additive opacity is a constant everywhere for all time. Traditionally,
we have used a value of 0.1 cm$^2$ g$^{-1}$ in our studies of Type Ia
light curves, but the best value will depend on SN type. A
comparison of SN Ib models calculated using CMFGEN and KEPLER
\citep{Ertl2020} suggested a value of 0.03 cm$^2$ g$^{-1}$ and we
regard this as a lower limit for the average. The actual value varies with location and time in a complicated way. Here we
adopt $\kappa_a = 0.1$ cm$^2$ g$^{-1}$ for the results presented in
this paper.

KEPLER carries a nuclear network of 19 species \citep{Weaver1978}
which do not perfectly correspond to the species in the
initial models provided.  Care was taken to translate the $^{56}$Ni
and $^{56}$Co abundances given to the zero-age $^{56}$Ni and
stable iron mass fractions which are used by KEPLER for energy
generation. The species $^{12}$C and $^{16}$O were mapped without
change. Other species such as $^{20}$Ne, $^{28}$Si, and so on were slightly
augmented where necessary by adding in nearby odd-Z abundances; that is, $^{20}$Ne included $^{20}$Ne and $^{23}$Na, $^{24}$Mg included
$^{24}$Mg and $^{27}$Al, and so on. As KEPLER does not compute spectra,
this should have negligible consequences.

The four models were mapped into KEPLER. The total mass, kinetic energy,
and $^{56}$Ni mass were preserved to 0.05\%. Because KEPLER is not a
special relativistic code, zones with velocities greater than $0.1c$
were trimmed from the input. High-velocity zones would also cause
difficulty with editing the luminosity at late times if the light-crossing time ceased to be negligible. Removing this high-velocity
material decreased the kinetic energy of all models by about 2\%. This
should have negligible effects on the light curve.
The zoning of the DDC models was
relatively coarse  by traditional KEPLER standards. After trimming the high-velocity zones, only 80 zones
remained. The zoning of the toy models was better with roughly 600
zones remaining. Rezoning was not carried out. For a light curve with
no nuclear burning, the zoning is adequate. All calculations took at
most a few minutes on a laptop.

\subsection{SEDONA}\label{sect:sedona}

SEDONA is a time-dependent, multi-frequency Monte Carlo radiative
transfer code originally developed to study SN light curves,
spectra, and polarisation \citep{Kasen2006}.
For this comparison study, the gas properties were tracked on 1D
spherical Lagrangian zones and are evolved under the assumption of
homologous expansion.

The radioactive decays of \nifs\ and \cofs\ were tracked and used to
source $\gamma$-ray packets. For the DDC10 model, radioactive decays
of $^{48}$Cr and $^{48}$V were also included in addition to the
\nifs\ decay chain. More detailed radioactive decay networks can be
implemented in SEDONA, but only the previously mentioned radioactive-decay processes were used in this comparison study.

 The $\gamma$-ray packets were transported using the Monte Carlo
 procedure subject to simplified treatments of bound-free absorption
 and Compton down-scattering. The $\gamma$-ray interactions were
 treated as heating terms that entered into the thermal heating
 balance that sourced a population of thermal photon packets.

The temperature in each shell was computed assuming radiative
equilibrium. In more detail, the LTE assumption let us set the gas
emissivity to the Planck function multiplied by the frequency-dependent line
opacity. The temperature was then iteratively adjusted until the
frequency-integrated emissivity was equal to the total radiative energy
from both $\gamma$-rays and re-emitted photon packets absorbed over
the previous particle propagation step, as measured in the co-moving
frame of the fluid.

For most of its published applications, SEDONA assumes LTE in order to
compute opacities, although non-LTE capabilities have been developed
and were  recently applied \citep{Roth2016,Shen2021}. For this comparison
study, only the LTE opacity mode was used in order to compute the
excitation and ionisation states of the gas. More precisely, this
means that ionisation fractions for each element were computed by
simultaneously solving the Saha equation using a local temperature,
and the charge conservation equation across all elements and isotopes
present.  Meanwhile, the bound-electron level populations within each
ionisation stage were set by the Boltzmann factors given by the local
temperature.

Photon packets were transported in three dimensions as a direct
solution to the time-dependent radiative transfer equation. For all
interactions with the gas, the packets were mapped to the 1D
Lagrangian zones.

Once the level populations had been computed, the bound-bound opacity was
computed using the expansion opacity formalism, as described by
\citet{Eastman1993} and \citet{Kasen2006}. For the toy01 and toy06
models, we only included the bound-bound opacities and electron scattering
opacity, using the Thomson cross section. For DDC10,
simplified bound-free and free-free opacities were also included in
addition to the previously mentioned opacities, although they did not
have a noticeable effect during the early stages of the explosion, which is what
we wish to compare using these calculations.

A thermalisation parameter $\epsilon$ of 1.0 was used for the toy01
and toy06 models. This means that all photon packets (other than
$\gamma$-ray packets) that were absorbed were immediately thermalised,
so that each absorption event was followed by re-emission of a photon
packet with a frequency sampled from the thermal emissivity. For the
DDC10 model, $\epsilon$ was set to 0.8, so in that case 20\% of
absorbed photon packets (not including the $\gamma$-ray packets) were
coherently scattered instead of having their frequency re-sampled.

The atomic data used for the bound-bound transitions were taken from
Kurucz CD 1. This is a larger set of atomic data than CD 23. The
details are described in Appendix~\ref{sect:atomdata_sedona}.

For this comparison study, the co-moving frequency grid for the
thermalised photons (i.e. not the $\gamma$-rays) used 17664 bins, with
equal logarithmic spacing, ranging from $10^{14}$ Hz to $2 \times
10^{16}$ Hz. The output spectra used 1044 frequency bins over the
range $1.1 \times 10^{14}$ Hz to $2 \times 10^{16}$ Hz
($\sim150$\,\AA\ to $\sim 2.7$\,$\mu$m). The time steps for the
homologous expansion and radioactive decay began at approximately a
few hours at the start of the calculation, and grew to no longer than
1 day, with a maximum time-step growth rate of 10\%. One million
$\gamma$-ray packets were emitted at each time step. As these $\gamma$
rays deposited their energy, their packets were discarded, while one
thermally sourced photon packet was emitted for each discarded
$\gamma$-ray packet. With all of the settings described above, a
calculation run to 100 days post-explosion required about 10 CPU
hours, and could be efficiently performed in parallel across several
hundred processors, reducing the elapsed wall time to less than one
hour.

\subsection{STELLA}\label{sect:stella}

The multi-group radiation hydrodynamics code STELLA
\citep{Blinnikov1993,Blinnikov1998,Blinnikov2006} is capable of
computing the evolution of the radiation field coupled to the
hydrodynamics, as well as the bolometric light curve, spectral energy
distribution, and resulting broad-band magnitudes and colours.  Therefore,
STELLA does not require the condition of homologous expansion and is
capable of treating shock propagation and any dynamical processes
taking place in the SN ejecta.

Energy deposition from \nifs\ and \cofs\ radioactive decay is treated
in a one-group diffusion approximation with an absorption opacity of
0.05 $Y_\mathrm{e}$\,cm$^{\,2}$\,g$^{\,-1}$ according to
\cite{Swartz1995}. The energy of positrons is thermalised locally.

The ionisation and level populations of a limited set of species (H,
He, C, N, O, Ne, Na, Mg, Al, Si, S, Ar, Ca, stable Fe, stable Co, and
stable Ni) is treated in the LTE approximation.
Radiation is not treated in equilibrium with the matter.  The colour
temperature is estimated as a black body temperature via the
least-squares method.

The opacity includes photoionisation, free–free absorption, and
electron scattering processes assuming LTE for the plasma and line
interactions. Radioactive \nifs\ and \cofs\ contribute to the stable
Fe when the line opacity is calculated.  The expansion opacity is
calculated according to \citep{Eastman1993}. The thermalisation
parameter for the line opacity treatment is set to 0.9
\citep{Kozyreva2020}. The line opacity is calculated using a data base
of 153\,441 spectral lines partly from \citet{Kurucz1995},
\cite{Verner1995}, and \cite{Verner1996_res}.

The spectral energy distribution is computed in the wavelength range
from 1\,\AA{} to 50\,000\,\AA{}. The frequency range is divided into
129 bins with equal logarithmic spacing, in which the radiative
transfer equations are solved at every time step. For the toy06 model,
a higher resolution simulation with 629 frequency bins was run, which
is the version used in the following sections for this model.  The
overall opacity might be slightly underestimated in the simulations
with 629 frequency bins compared to simulations involving 129 bins,
because the expansion opacity is calculated for the lines in the given
bin and is not extended to another bin even if the velocity gradient
is very high. The final (pseudo-)bolometric light curve is obtained by
integrating the spectra over frequency.

To avoid numerical artefacts, ejecta models closer in time to the
explosion were used (all are included in the data repository): 1\,h
post explosion for both toy models, and $\sim29$\,s and $\sim63$\,s
post explosion for models DDC10 and DDC25, respectively. The toy
models were computed on a lower-resolution grid (with 202 zones),
while the DDC models were computed at a higher resolution (with 399
zones). The typical runtime on a single processor is a few hours at
most.

\subsection{SUMO}\label{sect:sumo}

SUMO \citep{Jerkstrandthesis2011, Jerkstrand2011,Jerkstrand2012} is a
homologous non-LTE code with radiative transfer. It is specialised to
calculate spectra and light curves in the post-peak phases of the
SN. The code is written in Fortran and is parallelised with
MPI.

Gamma-ray transfer is done by ray tracing using a grey opacity of
$0.06 Y_e$\,cm$^2$\,g$^{-1}$. Positrons can be transferred (using an
effective opacity of 8.5 times the $\gamma$-ray one) but here they were
assumed to be locally trapped. The cascade of non-thermal electrons
following the scattering of gamma rays and positrons is solved for
with the Spencer-Fano method \citep{Kozma1992}.

Zone temperatures are solved from the first law of thermodynamics
considering heating by non-thermal processes and photoionisation, and
cooling by net collisional line excitation, free-free emission, and
recombination.  The temperature is solved either in steady state
(heating = cooling) \citep{Jerkstrand2012} or time-dependently
\citep{Pognan2022}. Here, the steady-state mode was used.

The rate equations are solved by considering spontaneous radiative
decay with Sobolev escape probabilities (assuming homology) modified
to include continuum processes and strong line overlaps, thermal and
non-thermal collisions (excitations and ionisations),
photoexcitation and de-excitation, photoionisation, and
recombination. Options exist to also include charge transfer processes
and molecular chemistry \citep{Liljegren2020}, but these were not
considered here. The rate equations can be solved either in steady
state (inflow = outflow) \citep{Jerkstrand2012} or time dependently
\citep{Pognan2022}. Here, the steady-state mode was used.

Radiative transfer is calculated with a Monte Carlo simulation that is
iterated with the solvers for temperature,
ionisation, and excitation. During the transfer, photon packets can
experience electron scattering, bound-free and free-free absorption,
and line absorption. The line transfer is resolved line by line (with
Sobolev formalism) rather than using an expansion opacity
formalism. Photoexcitation is either fully coupled to the non-LTE
solutions (for low- and mid-lying levels), or decoupled (for high-lying
levels), instead giving a fluorescence cascade on the spot. The
radiation field is computed in steady state.

The atomic data come from a variety of sources, mostly described in
\cite{Jerkstrand2011} and \cite{Jerkstrand2012}. Model atoms have
LS-states resolved, that is they do not use super levels. In models computed
up to 2022 (including the ones here), ions up to and including doubly
ionised were included, although higher ions are now being included
\citep{Pognan2022}. For the toy models computed here, the model atoms
are Fe\one\ (496 levels), Fe\two\ (578 levels), Fe\three\ (600
levels), Ni\one\ (136 levels), Ni\two\ (500 levels), Ni\three\ (8
levels), Co\one\ (317 levels), Co\two\ (108 levels), Co\three\ (306
levels), Si\one\ (494 levels), Si\two\ (77 levels), Si\three\ (2
levels), S\one\ (123 levels), S\two\ (5 levels), S\three\ (6 levels),
Ca\one\ (198 levels), Ca\two\ (69 levels), and Ca\three\ (1
level)\footnote{For Ni\three\, Si\three, S\two\ and Ca\three\ the low
number of levels is sufficient as the next state energies are at
52\,152, 52\,853, 79\,395 and 203\,373 cm$^{-1}$, respectively.}.

For the runs here, the ejecta were resampled to $\Delta v =
500$\,\kms, and truncated at $v = 30000$\,\kms. The radiative transfer
was computed on a wavelength grid from 400\,\AA\ to 25000\,\AA, with a
logarithmic resolution $\mathrm{d}\lambda/\lambda = 10^{-3}$. Typical
wall-clock runtimes for a single epoch are a few hours on a typical 128
core setup.

\subsection{SuperNu}\label{sect:supernu}

SuperNu is a multi-group LTE radiative-transfer code that employs
Implicit Monte Carlo (IMC) and Discrete Diffusion Monte Carlo (DDMC)
\citep[][]{Wollaeger2013,Wollaeger2014}.  IMC solves the thermal
radiative-transfer equations semi-implicitly by treating some
absorption and emission as instantaneous effective scattering
\citep[see e.g. ][]{Fleck1971}.  DDMC optimises IMC in optically thick
regions of space \citep[][]{Densmore2012} and ranges of wavelength
\citep[][]{Abdikamalov2012,Densmore2012} by replacing many low
mean-free-path scattering events with single leakage events.  SuperNu
can apply IMC and DDMC in both static and homologous,
semi-relativistically expanding atmospheres.  The code has been
verified by analytic and semi-analytic radiative transfer tests
\citep[][]{Wollaeger2013} and on the W7 model of SNe Ia
\citep[][]{Nomoto1984,Wollaeger2014}.

For the $\gamma$-ray transfer, SuperNu employs a constant absorption
opacity of 0.06 $Y_e$~cm$^2$\,g$^{-1}$, where $Y_e$ is ejecta gas
electron fraction, following the prescription
of~\citet[][]{Swartz1995}.  The $\gamma$-ray packets in SuperNu are
not directly converted to optical packets, but instead are used to
tally the total $\gamma$-ray energy deposition per spatial cell.  The
deposition energy values are then added to the thermal source for
optical packets.

The ejecta gas temperature is calculated using the standard IMC
semi-implicit linearisation~\citep[][]{Fleck1971} of the comoving
internal energy equation~\citep[][]{Wollaeger2013}: internal energy is
recast to gas temperature using the standard relation $\partial e =
c_v\partial T$, where $e$ is internal energy, $c_v$ is heat capacity
at constant volume, and $T$ is gas temperature.  Due to the IMC time
linearisation, energy deposited from gamma-rays and (locally) from
beta particles appears simultaneously in both the comoving internal
energy equation and the radiation equation.

Ionisation and excitation are both treated with Saha-Boltzmann statistics
evaluated at the gas density and temperature at the beginning of
the time step.  The multi-element system is solved iteratively by
converging the free electron number.  The resulting population
densities are then used to calculate opacities.  The radiation field
is represented by fully time-dependent Monte Carlo packets, which are
sourced from the LTE emissivity.  The radiation field is not constrained
to be in equilibrium with the gas, and so in general the system is
`two-temperature'.

Opacity is discretised into groups via direct integration over
co-moving wavelength (but see \citealt{Fontes2020} for a study with
SuperNu using expansion opacity).  Opacity in SuperNu includes
free-free \citep{Sutherland1998}, and bound-free
\citep{Verner1995,Verner1996} processes, as well as the bound-bound
opacities from the Kurucz line
lists\footnote{\url{http://kurucz.harvard.edu/atoms.html}}, and the
standard elastic Thomson scattering opacity.  Weak lines in the Kurucz
data set are omitted from the SuperNu line list where the opacity is
dominated by stronger lines.  The total number of available lines for
the present simulations is 591\,288.  This list was motivated by
studies using the PHOENIX code in the work of \cite{vanRossum2012},
using the full line list (>10$^{7}$ lines) as a benchmark.  For the
simulations in this work, opacity is computed in 1000 logarithmic
wavelength groups from 100 to 32000\,\AA.

Each simulation presented for SuperNu uses 4\,194\,304 source packets
per time step, with maximum active packet populations of between 100 and
200 million.  The wavelength group structure that the packets are
tracked through is the same as the opacity group structure (though for
the simulations here, the wavelength bounds for the flux tallies are
1000 and 32,000\,\AA).  The DDC and toy models use 115 and 202
velocity space cells, respectively.  For the time grid, the DDC and
toy models use 200 logarithmically increasing time steps out to
day 80, from about day 1 (DDC models) or day 2 (toy models)
post explosion.

SuperNu has MPI+OpenMP parallelisation.  The SuperNu simulations of
the toy01, toy06, DDC10, and DDC25 models presented here cost 190,
320, 398, and 400 core-hours, respectively, each using 16 MPI ranks
and 8 OpenMP threads per rank.

\subsection{TARDIS}\label{sect:tardis}

TARDIS is an open-source steady-state 1D radiative-transfer code that
uses indivisible energy packets as its transport quanta following the
methods in
\citet{Abbot1985}, \cite{lucy1993}, \cite{Mazzali1993},
\cite{Lucy2002}, \cite{Lucy2003}, and \cite{Lucy2005}. \citet{kerzendorf2014}
describes the initial version of the code which was primarily used to
model SNe Ia. Subsequently, the code has been significantly enhanced
to include a non-thermal approximation treatment for helium
\citep{boyle2017} and the continuum processes and relativity
treatments required for hydrogen-rich SNe \citep{vogl2019}.  TARDIS
has been continuously enhanced since then \citep[see e.g. ][similarly
  gamma-ray energy deposition is a new module of TARDIS, but has not
  been used for the models in this work]{kerzendorf2022}.  Full
documentation and an extended physics walk-through for TARDIS can be
found online\footnote{\url{https://tardis-sn.github.io/tardis}}.

TARDIS assumes a steady-state homologously expanding
SN envelope and injects Monte Carlo packets ---randomly sampled
from a given distribution (by default a black-body)--- from an inner
boundary. The code supports bound-bound, bound-free, free-free, and
Thomson opacities with several redistribution schemes from simple
scattering to a macro-atom \citep{Lucy2002, Lucy2003}. Summary packet
statistics are used to estimate radiation field quantities
(temperature, dilution factors, mean intensities, heating- and
photo-ionisation rates, and line source functions). The estimated
quantities are used to calculate the ionisation and excitation populations
in steady state with a choice between LTE and several formulations of
non-LTE (nebular approximation of \citealt{Abbot1985} for ionisation,
and dilute-LTE and full non-LTE for excitation). TARDIS then calculates
Sobolev optical depths for line interaction, and opacities for
continuum processes. These values are then used in the subsequent
Monte Carlo step, which produces summary statistics for updating the
opacities. The other convergence criterion is a match of the
integrated packet output luminosity and the requested luminosity,
which is achieved through iterative adjustments of the inner
temperature.

To handle atomic data, the TARDIS collaboration has developed an
additional package named \textsc{Carsus} \citep[available at
  \url{https://github.com/tardis-sn/carsus};][]{Passaro2019_4062427}
with documentation available
online \footnote{\url{https://tardis-sn.github.io/carsus}}.
\textsc{Carsus} can read atomic data (masses, ionisation energies,
levels, lines, photoionisation cross sections, and collisional cross
sections) from NIST \citep{kramida_nist_1999}, Chianti
\citep{Dere_2019, Dere_1997}, CMFGEN \citep{Hillier1998_cmfgen,
  Dessart2010}, and Kurucz \citep{Kurucz2009_ATD}. For the models in
this paper, we used Kurucz CD 23 as the source of atomic data (see
Appendix~\ref{sect:atomdata_tardis} for the full description).

The setup files for TARDIS that are used in this work are available
online\footnote{\url{https://github.com/tardis-sn/tardis-setups/tree/master/2022/sn_radtrans_compare}}
using an atomic data set from Kurucz CD 23. In the setup used for the
code comparison work, we run TARDIS in a mode that self-consistently
finds a temperature stratification given an inner boundary velocity
and output luminosity. We used the mean bolometric luminosity of the
other comparison codes for our output luminosity. The structure of the
toy01 and toy06 models was set at 520 shells between 9000 and 35000 km
s$^{-1}$. The DDC models were truncated to 40 shells between 9000 and
35000 km s$^{-1}$.  The inner boundary velocity is found by iterating
until the dilution factor is close to 0.5 in the innermost zone.
Table~\ref{tab:tardis_lum_vel} shows the inner boundary velocity and
requested output luminosity for the different epochs.

\begin{table}
\centering
\footnotesize
\caption{Velocity and temperature at the inner boundary given the requested output luminosity for the TARDIS calculation of the toy06 model.}
\label{tab:tardis_lum_vel}
\begin{tabular}{cccc}
\hline
\multicolumn{1}{c}{Time} & $\log_{10} L_\mathrm{req}$ & \multicolumn{1}{c}{Velocity at IB} & \multicolumn{1}{c}{Temperature at IB} \\
\multicolumn{1}{c}{(days)} & (erg\,s$^{-1}$) & \multicolumn{1}{c}{(\kms)} & \multicolumn{1}{c}{(K)} \\
\hline
5  & 42.48 & 20500 &   8805 \\
10 & 42.95 & 17000 &   8870 \\
15 & 43.04 & 10000 &  11525 \\
20 & 43.00 &  5500 &  15513 \\
\hline
\end{tabular}
\flushleft
\textbf{Notes:} $L_\mathrm{req}$ = requested output luminosity; IB = inner boundary.
\end{table}

We used the \texttt{nebular} approximation for ionisation and the
\texttt{dilute-lte} approximation for excitation. We used
$5\times10^5$ packets for estimating our radiation field and 30
iterations for convergence. In the final iteration, we estimated the
source functions with $5\times10^5$ packets and then used the formal
integral to synthesise a spectrum. For the comparison, we use line
opacities and Thomson opacities with the \texttt{macroatom}
interaction scheme. The resolution of the output spectra was uniform
from 500 to 20000 \AA\ with 2 \AA\ bin width. The models were run on
one CPU with runtimes of less than one hour.

\subsection{URILIGHT}\label{sect:urilight}

URILIGHT is a time-dependent Monte-Carlo code written in Fortran 90 by
Yoni Elbaz based on the approximations that are used in SEDONA (see
Sect.~\ref{sect:sedona} above, \citealt{Kasen2006} and references
therein), in particular assuming homologous expansion. A detailed
description of this program and previous comparisons to other
published radiative-transfer codes for several benchmark problems are
presented in \cite{Wygoda2019}. The code is publicly available and can
be downloaded from
\url{https://www.dropbox.com/sh/kyg1z1xwi0298ru/AAAqzUMbr6AkoVfkSVIYChTLa?dl=0}.

Energy deposition resulting from the decay of radioactive isotopes is
calculated by a Monte-Carlo solution of the $\gamma$-ray transport,
for which interaction with matter is included through Compton
scattering and photoelectric absorption. For the calculations in this
work, only $^{56}$Ni and $^{56}$Co decay were included.
The temperature is iteratively solved for in each cell by requiring
that the total emissivity be equal to the total absorbed energy.
LTE is assumed for calculating the ionisation and excitation states:
ionisation is obtained by solving the Saha equation, and excitation
levels are set by the Boltzmann-distribution.

Opacities include bound-bound and free-free absorption and Thomson
scattering off free electrons. The atomic line data for the
bound-bound transitions, which constitutes the main and most important
source of opacity in \sneia, are taken from the extended set of lines
by Kurucz\footnote{CDs 1 and 23 from
\url{http://kurucz.harvard.edu/cdroms.html}}. Following a bound-bound
interaction, most photons (a fraction $\epsilon$, which is a global
parameter of the simulation; see \citealt{Kasen2006}) are thermalised
and re-emitted at a different wavelength with a distribution set by
the emissivity. The rest of the photons (fraction $1-\epsilon$) are
re-emitted at the same wavelength (within the line width). As in
\cite{Wygoda2019}, we used $\epsilon=0.8$ in the runs performed
here.

The runs here use 162 spatial cells for the toy01 and toy06 models
and 115 spatial cells for the DDC10 and DDC25 models. All models are
run with 128 time steps logarithmically spaced between 2 and 210
days, and a uniform spectrum resolution of 10\,\AA\ between 100 and
30000\,\AA. Each run typically took 30 hours on a single core, or
approximately one hour when parallelised.

%%% Table with code description
\begin{table*}
\centering
\footnotesize
\caption{Code outputs and computed models.}\label{tab:codedescr}
\begin{tabular}{lcccccccccccc}
\hline
Code        & RT        & Bolometric      & Spectrum &  Early       & Nebular       & \multicolumn{4}{c}{Computed Models} \\  
Name        & Method    & Flux            & or SED   &  Times       & Times         & toy06  & toy01  & DDC10  & DDC25  \\
\hline
ARTIS       & MC        & calculated      & \cmark   & \cmark       & \cmark        & \cmark & \xmark & \cmark & \xmark \\
CMFGEN      & RTE-CMF   & calculated      & \cmark   & \cmark       & \cmark        & \cmark & \cmark & \cmark & \cmark \\
CRAB        & RH-1G     & calculated      & \xmark   & \cmark       & \xmark        & \cmark & \xmark & \xmark & \xmark \\
KEPLER      & FLD       & calculated      & \xmark   & \cmark       & \xmark        & \cmark & \cmark & \cmark & \cmark \\
SEDONA      & MC        & calculated      & \cmark   & \cmark       & \xmark        & \cmark & \cmark & \cmark & \xmark \\
STELLA      & RH-MG     & calculated      & \cmark   & \cmark       & \xmark        & \cmark & \cmark & \cmark & \cmark \\
SUMO        & MC        & calculated      & \cmark   & \xmark       & \cmark        & \cmark & \cmark & \xmark & \xmark \\
SuperNu     & MC        & calculated      & \cmark   & \cmark       & \xmark        & \cmark & \cmark & \cmark & \cmark \\
TARDIS      & MC        & input           & \cmark   & \cmark       & \xmark        & \cmark & \cmark & \cmark & \cmark \\
URILIGHT    & MC        & calculated      & \cmark   & \cmark       & \xmark        & \cmark & \cmark & \cmark & \cmark \\
\hline
\end{tabular}
\flushleft
\textbf{Notes:} RT Method gives the numerical method used to solve the radiative-transfer equation (see Table~\ref{tab:codephys}).
\end{table*}

%%%%%%%%%%%%%%%%%%%%%%%%%%%%%%%%%%%%%%%
% data repository
%%%%%%%%%%%%%%%%%%%%%%%%%%%%%%%%%%%%%%%

\section{Data repository of test models and standardised outputs}\label{sect:outputs}

The ejecta models and output files from the RT simulations of the
different codes are provided in a new data repository, which is
publicly available and can be accessed at
\url{https://github.com/sn-rad-trans/data1}.

Descriptions of the files available in the repository are
provided below, including ejecta model files
(Sect.~\ref{sect:outputs:model_formats}), output files
(Sect.~\ref{sect:outputs:result_formats}), and Python codes that were
used to create the analytic toy model ejecta and codes for reading the
output files (Sect.~\ref{sect:outputs:codes}).

\subsection{Ejecta model files}\label{sect:outputs:model_formats}

As described in Sect.~\ref{sect:models}, the code comparison is
performed using four SN Ia models (main parameters in Table
\ref{tab:models}).  The RT input files that were distributed among the
groups are provided in the repository, including two toy-model files
\texttt{snia\_toy01\_2d.dat} and \texttt{snia\_toy06\_2d.dat} and the two
delayed-detonation model files \texttt{DDC10\_0.976d.dat} and
\texttt{DDC25\_1.300d.dat}.

\subsubsection{Toy model files}

The two toy-model files, \texttt{snia\_toy01\_2d.dat} and
\texttt{snia\_toy06\_2d.dat}, represent a snapshot of the ejecta at 2.0
days post explosion and include 807 shells (rows) with the following
21 columns\footnote{For the STELLA runs, the files
\texttt{snia\_toy01\_1h\_lowres.dat} and
\texttt{snia\_toy06\_1h\_lowres.dat} represent a snapshot of the
ejecta at 1\,h post explosion and include 202 shells (rows) with the
same structure.}:

\begin{enumerate}[label=(\arabic*), align=left]
\item Shell index (1-807)
\item Velocity at shell centre (\kms) 
\item Shell mass (\msun)
\item Lagrangian mass coordinate at the outer shell boundary (\msun)
\item Pre-decayed ($t=0$) stable IGE mass fraction (0 for all shells) 
\item Pre-decayed ($t=0$) $^{56}$Ni mass fraction 
\item IME mass fraction (of which 10\% is Ca, 35\% is S, and 55\% is Si by mass) 
\item Ti mass fraction (0 for all shells)
\item Unburnt C+O mass fraction (0 for all shells)
\item Radius at shell centre (cm) = velocity at shell centre $\times$ 2 days (homologous expansion)
\item Mean density over shell (\gcc), {not} density at shell centre
\item Temperature (K)
\item[(13)-(21)] Mass fractions of $^{56}$Ni, $^{56}$Co, $^{56}$Fe, Ca, S, Si, O, and C at 2 days post explosion
\end{enumerate}

\subsubsection{Delayed-detonation model files}

The two delayed-detonation model files \texttt{DDC10\_0.976d.dat} and
\texttt{DDC25\_1.300d.dat} represent a snapshot of the ejecta at 0.976
days and 1.3 days, respectively, with 115 shells with the following 50
columns\footnote{For the STELLA runs, the files
\texttt{DDC10\_29.29s\_highres.dat} and
\texttt{DDC25\_62.60s\_highres.dat} represent a snapshot of the
ejecta at $\sim29$\,s and $\sim63$\,s post explosion, respectively,
with 399 shells. The structure is the same but includes four
additional columns giving the elemental mass fractions of H, He, N,
and P.}:

\begin{enumerate}[label=(\arabic*), align=left]
    \item Velocity at shell centre (\kms)
    \item Radius at shell centre (cm) = velocity at shell centre
      $\times$ the time post explosion
    \item Shell volume (cm$^3$)
    \item Density at zone centre (\gcc), {not} mean density
      over shell
    \item Shell mass estimate (g) = shell volume $\times$ density at
      zone centre
    \item Temperature (K)
    \item[(7)-(26)] Elemental mass fractions at snapshot time of C, O,
      Ne, Na, Mg, Al, Si, S, Cl, Ar, K, Ca, Sc, Ti, V, Cr, Mn, Fe, Co,
      and Ni
    \item[(27)-(50)] isotopic mass fractions of the following
      radioactive nuclei at snapshot time: $^{56}$Ni, $^{56}$Co,
      $^{57}$Ni, $^{57}$Co, $^{48}$Cr, $^{48}$V, $^{49}$Cr, $^{49}$V,
      $^{51}$Mn, $^{51}$Cr, $^{55}$Co, $^{55}$Fe, $^{37}$K, $^{37}$Ar,
      $^{52}$Fe, $^{52}$Mn, $^{44}$Ti, $^{44}$Sc, $^{41}$Ar,
      $^{42}$Ar, $^{42}$K, $^{43}$K, $^{47}$Sc, and $^{61}$Co
\end{enumerate}

\subsection{RT output files}\label{sect:outputs:result_formats}

For each simulation by a specific group, six output file types are
generated for each of the ejecta models.

\subsubsection{Output file names}\label{sect:outputs:output_file_names}

The name of each file has the following structure:

\begin{verbatim}
    <output type>_<model>_<code name>.txt
\end{verbatim}

\noindent
where 

\begin{itemize}

    \item \verb+<output type>+ represents the type of output
      (described below) and can take one of six values:
      \texttt{lbol\_edep}, \texttt{edep}, \texttt{phys},
      \texttt{ionfrac\_<element>} (where \texttt{<element>} is the
      name of a given element, e.g.  \texttt{ca} for calcium),
      \texttt{spectra}, and \texttt{wsynphot\_mags},\\

    \item \verb+<model>+ represents one of the models and can take one
      of four values: \texttt{toy06}, \texttt{toy01}, \texttt{ddc10},
      and \texttt{ddc25},\\
    
    \item \verb+<code name>+ represents the code name with an optional
      additional descriptor (useful to distinguish between different
      code settings when applied to a given model) and can take one of
      12 values: \texttt{artis}, \texttt{artisnebular},
      \texttt{cmfgen}, \texttt{crab}, \texttt{kepler},
      \texttt{sedona}, \texttt{stella}, \texttt{stella\_fr600} (for
      the STELLA runs for the toy06 models that use 629 frequency bins
      instead of the default 129), \texttt{supernu}, \texttt{sumo},
      \texttt{tardis}, and \texttt{urilight}.
    
\end{itemize} 

In principle, there are 12 code names $\times$ 4 ejecta models $\times$
5 output files (excluding the \texttt{ionfrac\_<element>} files) = 240
files and an additional 12 code names $\times$ (2 toy models $\times$
6 elements + 2 DDC models $\times$ 20 elements) = 624
\texttt{ionfrac\_<element>} files. This results in a total of 864
files, although in practice not all files are available for various
reasons: (a) a code was not applied to a given model (it was agreed
that all groups should at least compute the toy06 model, but the other
three were considered optional); (b) a given code cannot produce the
specified output (e.g. SEDs for grey codes); (c) a given code does not
provide the desired output quantities by default (i.e. modification of
the source code would be necessary). Table~\ref{tab:codedescr}
summarises the outputs and computed models for each code.

\subsubsection{Six output file types}\label{sect:outputs:output_file_types}

The six types of output files include:

\begin{enumerate}

    \item \textbf{Pseudobolometric (UVOIR) luminosity and global energy deposition as a function of time.}\\
    
    \textbf{File name:} \verb+lbol_edep_<model>_<code name>.txt+\\

    \textbf{Header}: after some optional comment lines, the following
    two lines give the number of epochs (\texttt{NTIMES}, 100 in this
    example) and the column headings:
    
    \begin{verbatim}
    #<optional comment lines>
    #NTIMES: 100
    #time[d] Lbol[erg/s] Edep[erg/s]
    \end{verbatim}
    \vspace*{-1em}
    
    \textbf{Contents}: \texttt{NTIMES} rows with the following three columns:
    
    \begin{enumerate}[label=(\arabic*), align=left]
    \item Time since explosion in days
    \item Pseudobolometric (UVOIR) luminosity in erg s$^{-1}$
    \item Global energy deposition by $\gamma$ rays and positrons in erg s$^{-1}$
    \end{enumerate}

    \vspace*{1em}
    \item \textbf{Energy deposition}\\
    
    \textbf{File name:} \verb+edep_<model>_<code name>.txt+\\

    \textbf{Header}: after some optional comment lines, the following
    four lines give the number of epochs (\texttt{NTIMES}, 100 in this
    example), the number of cells (\texttt{NVEL}, 200 in this
    example), the list of epochs in days with all \texttt{NTIMES}
    values (the `...' should correspond to actual values), and finally
    the column headings (here the `...' can be used as is):
    
    \begin{verbatim}
    #<optional comment lines>
    #NTIMES: 100
    #NVEL: 200
    #TIMES[d]: 2.0 3.0 ... 100.0
    #vel_mid[km/s] Edep_t0[erg/s/cm^3] 
     Edep_t1[erg/s/cm^3] ... Edep_tn[erg/s/cm^3]
    \end{verbatim}
    \vspace*{-1em}
    
    \textbf{Contents}: \texttt{NVEL} rows with the following \texttt{NTIMES}+1 columns (101 in this example):
    
    \begin{enumerate}[label=(\arabic*), align=left]
    \item Velocity at the centre of each cell in \kms
    \item[(2)-(101)] Total $\gamma$-ray + positron energy deposition
      rate at each of the \texttt{NTIMES} epochs within the
      corresponding cell in erg s$^{-1}$ cm$^{-3}$
    \end{enumerate}

    \vspace*{1em}
    \item \textbf{Physical conditions}\\
    
    \textbf{File name:} \verb+phys_<model>_<code name>.txt+\\

    \textbf{Header}: after some optional comment lines, the following
    three lines give the number of epochs (\texttt{NTIMES}, 100 in
    this example), the list of epochs with all \texttt{NTIMES} values
    (the `...' should correspond to actual values), and finally one
    empty comment line:
    
    \begin{verbatim}
    #<optional comment lines>
    #NTIMES: 100
    #TIMES[d]: 2.0 3.0 ... 100.0
    #
    \end{verbatim}
    \vspace*{-1em}
    
    \textbf{Contents}: \texttt{NTIMES} blocks (one for each epoch),
    each containing a block header with three lines giving the epoch
    (in days, 2.0\,d in this example), the number of cells saved for
    this epoch (\texttt{NVEL}, 200 in this example), and finally the
    column headings:
    
    \begin{verbatim}
    #TIME: 2.0
    #NVEL: 200
    #vel_mid[km/s] temp[K] rho[gcc] ne[/cm^3]
     natom[/cm^3] 
    \end{verbatim}
    \vspace*{-1em}

    Each block content consists of \texttt{NVEL} rows with the following five columns:

    \begin{enumerate}[label=(\arabic*), align=left]
    \item Velocity at the centre of each cell in \kms
    \item Temperature in K
    \item Density in \gcc
    \item Free electron density in cm$^{-3}$
    \item Total atom density in cm$^{-3}$
    \end{enumerate}

    \vspace*{1em}
    \item \textbf{Ionisation fraction (one file per element)}\\
    
    \textbf{File name:}\\
    \verb+ionfrac_<element>_<model>_<code name>.txt+\\

    \textbf{Header}: after some optional comment lines, the following
    four lines give the number of epochs (\texttt{NTIMES}, 100 in this
    example), the number of ionisation stages (\texttt{NSTAGES}, 6 in
    this example, starting at neutral and up to $\texttt{NSTAGES}-1$
    times ionised), the list of epochs with all \texttt{NTIMES} values
    (the `...' should correspond to actual values), and finally one
    empty comment line:
    
    \begin{verbatim}
    #<optional comment lines>
    #NTIMES: 100
    #NSTAGES: 6
    #TIMES[d]: 2.0 3.0 ... 100.0
    #
    \end{verbatim}
    \vspace*{-1em}
    
    \textbf{Contents}: \texttt{NTIMES} blocks (one for each epoch),
    each containing a block header with three lines giving the epoch
    (in days, 2.0\,d in this example), the number of cells saved for
    this epoch (\texttt{NVEL}, 200 in this example), and finally the
    column headings (where we consider the element Fe in this
    example):
    
    \begin{verbatim}
    #TIME: 2.0
    #NVEL: 200
    #vel_mid[km/s] fe0 fe1 fe2 fe3 fe4 fe5 
    \end{verbatim}
    \vspace*{-1em}

    Each block content consists of \texttt{NVEL} rows with the
    following \texttt{NSTAGES}+1 columns (7 in this example):

    \begin{enumerate}[label=(\arabic*), align=left]
    \item Velocity at the centre of each cell in \kms
    \item[(2)-(7)] Fraction of ions (dimensionless) in the
      corresponding cell (\texttt{fe0} = Fe\one, \texttt{fe1} =
      Fe\two\ etc.). The sum of the fractions in each of the
      \texttt{NVEL} rows is expected to be unity.
    \end{enumerate}

    \vspace*{1em}
    \item \textbf{Spectral sequence}\\
    
    \textbf{File name:} \verb+spectra_<model>_<code name>.txt+\\

    \textbf{Header}: after some optional comment lines, the following
    four lines give the number of epochs (\texttt{NTIMES}, 100 in this
    example), the number of wavelengths (\texttt{NWAVE}, 2000 in this
    example), the list of epochs in days with all \texttt{NTIMES}
    values (the `...' should correspond to actual values), and finally
    the column headings (here the `...' can be used as is):
    
    \begin{verbatim}
    #<optional comment lines>
    #NTIMES: 100
    #NWAVE: 2000
    #TIMES[d]: 2.0 3.0 ... 100.0
    #wavelength[Ang] flux_t0[erg/s/Ang]
     flux_t1[erg/s/Ang] ... flux_tn[erg/s/Ang]
    \end{verbatim}
    \vspace*{-1em}

    \textbf{Contents}: \texttt{NWAVE} rows with the following
    \texttt{NTIMES}+1 columns (101 in this example):
    
    \begin{enumerate}[label=(\arabic*), align=left]
    \item Wavelength in \AA
    \item[(2)-(101)] Fluxes at each of the \texttt{NTIMES} epochs at
      the corresponding wavelength in erg s$^{-1}$ \AA$^{-1}$
    \end{enumerate}

    \vspace*{1em}
    \item \textbf{Synthetic photometry}\\
    
    \textbf{File name:} \verb+wsynphot_mags_<model>_<code name>.txt+\\

    \textbf{Header}: after some optional comment lines, the following
    three lines give the number of epochs (\texttt{NTIMES}, 100 in
    this example), the number of photometric bands (\texttt{NBANDS}, 8
    in this example), and finally the column headings (giving each
    band name, $UBVRIJHK$ in our example):
    
    \begin{verbatim}
    #<optional comment lines>
    #NTIMES: 100
    #NBANDS: 8
    #time[d] U B V R I J H K
    \end{verbatim}
    \vspace*{-1em}
    
    \textbf{Contents}: \texttt{NTIMES} rows with the following \texttt{NBANDS}+1 columns (9 in our example):
    
    \begin{enumerate}[label=(\arabic*), align=left]
    \item Time since explosion in days
    \item[(2)-(9)] Absolute magnitudes in each band at the corresponding time 
    \end{enumerate}

\end{enumerate}

\subsection{Useful Python codes}\label{sect:outputs:codes}

All the figures presented in this paper are automatically generated
and accessible in a Python notebook \verb+all_plots.ipynb+ which is
available in the data repository. A separate Python notebook
\verb+photometry.ipynb+ in the \verb+code-comparison1/+ directory is
used to generate the synthetic photometry (\verb+wsynphot_mags+ files)
from the \verb+spectra+ files on the fly. Three further useful Python
codes can be used to:

\begin{enumerate}

    \item \textbf{Generate the toy models}\\
    
    \textbf{Location:}\\ \verb+data1/input_models/mk_snia_toy_model.py+\\
    
    \textbf{Calling syntax:}
    \begin{verbatim}
    python mk_snia_toy_model.py --highni
    python mk_snia_toy_model.py --lowni
    \end{verbatim}
    \vspace*{-1em}
    This code generates the toy models used in this paper, as
    explained in Sect.~\ref{sect:toymodels}. The upper command
    generates the toy06 model (\verb+snia_toy06_2d.dat+), while the lower
    line generates the toy01 model (\verb+snia_toy01_2d.dat+). Comments
    inside the code provide information on how to create new similar
    test models with different parameters.

    \vspace*{1em}
    \item \textbf{Read the input files}\\
    
    \textbf{Location:} \verb+data1/input_models/read_inputs.py+\\
    
    \textbf{Calling syntax:}
    \begin{verbatim}
    python read_inputs.py
    \end{verbatim}
    \vspace*{-1em}
    This code reads the input files of the test models and creates the
    following files corresponding to Figs.~\ref{fig:dens} and
    \ref{fig:composition} as well as Table~\ref{tab:models}:
    \verb+density_profile.pdf+ (Fig.~\ref{fig:dens}),
    \verb+composition_profile_<model>.pdf+ (where \texttt{<model>} is
    one of \texttt{toy06}, \texttt{toy01}, \texttt{ddc10}, and
    \texttt{ddc25}; Fig.~\ref{fig:composition}), and
    \texttt{models\_summary.tex} (Table~\ref{tab:models}). The code
    also includes the functions \texttt{read\_snia\_toy\_model()} and
    \texttt{read\_ddc\_model()} to read the files into Python
    variables.

    By default, the code is expected to be executed from within the
    \verb+data1/input_models/+ directory, although the path to the
    \texttt{data1/} directory can be specified using the
    \verb+--path2data+ option.

    \vspace*{1em}
    \item \textbf{Read the output files}\\
    
    \textbf{Location:} \verb+data1/read_outputs.py+\\
    
    \textbf{Calling syntax:}
    \begin{verbatim}
    python read_outputs.py file1.txt file2.txt
    python read_outputs.py /path/to/file*.txt
    \end{verbatim}
    \vspace*{-1em}
    This code reads the six output file types (see
    \ref{sect:outputs:output_file_types}) and produces corresponding
    plots. The code also includes the functions:
    \texttt{read\_lbol\_edep()}, \texttt{read\_spectra()},
    \texttt{read\_edep()}, \texttt{read\_phys()},
    \texttt{read\_ionfrac()}, and \texttt{read\_mags()} to read the
    output files into Python variables.
    
    The code can be executed from any directory because the full path of
    each file can be specified (wildcards are also accepted). When
    uploading new output files to the data repository users are
    required to ensure they match the expected format exactly. This
    can be achieved using the \verb+--checkformat+ option (and
    optionally the \verb+--noplot+ option to disable the plotting
    functionality when checking a large number of files).
    
\end{enumerate}

%%%%%%%%%%%%%%%%%%%%%%%%%%%%%%%%%%%%%%%
% Results
%%%%%%%%%%%%%%%%%%%%%%%%%%%%%%%%%%%%%%%

\section{Example results}\label{sect:results}

\begin{figure*}
\centering
 \includegraphics[width=0.85\hsize]{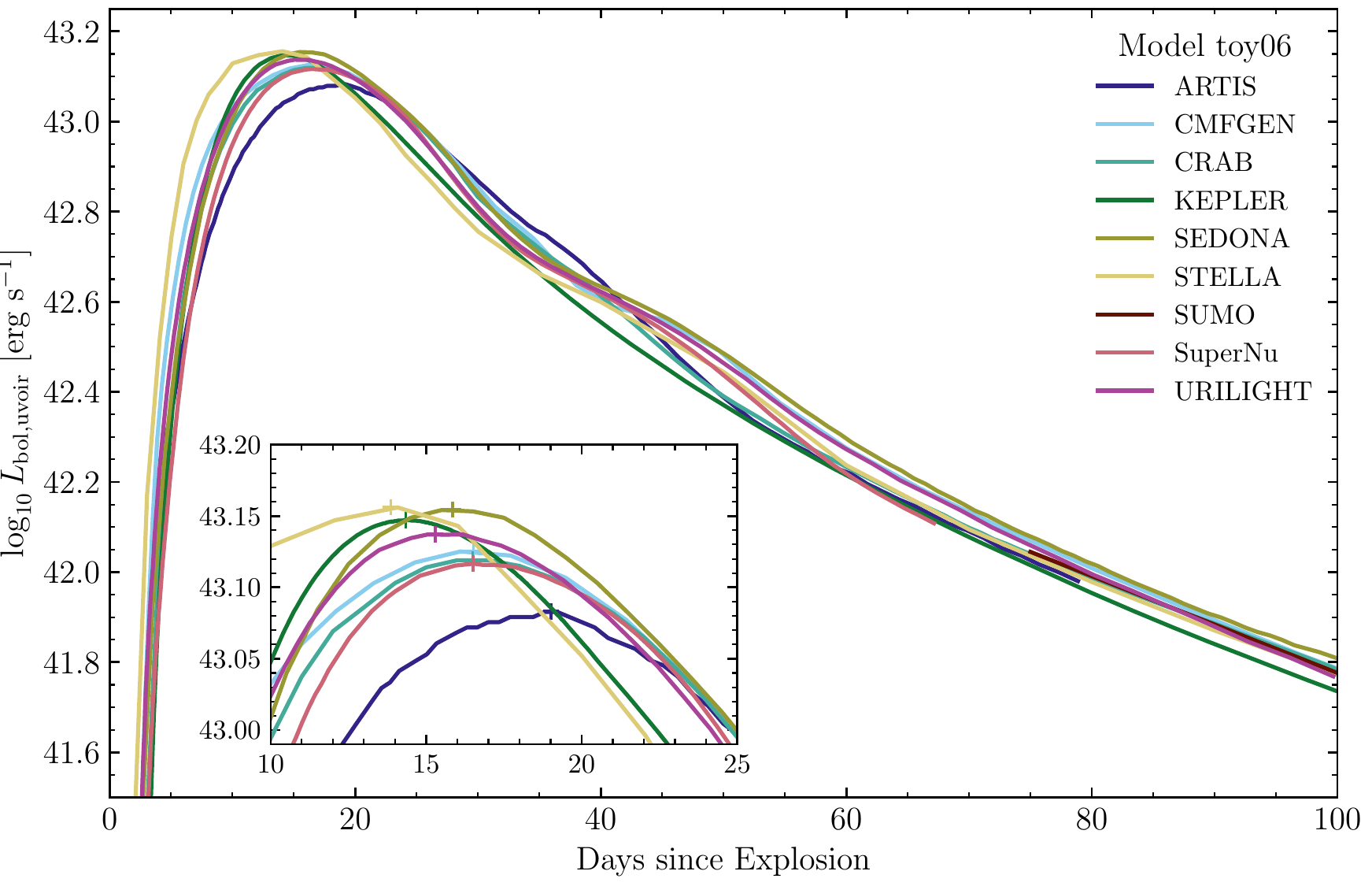}
   \caption{Pseudobolometric (UVOIR) light curves for model
     toy06. The inset shows a zoom into the maximum-light epoch (the
     estimated time of maximum light is indicated with a `+' sign).
   }
 \label{fig:lbol}
\end{figure*}

\begin{figure}
 \includegraphics[width=\hsize]{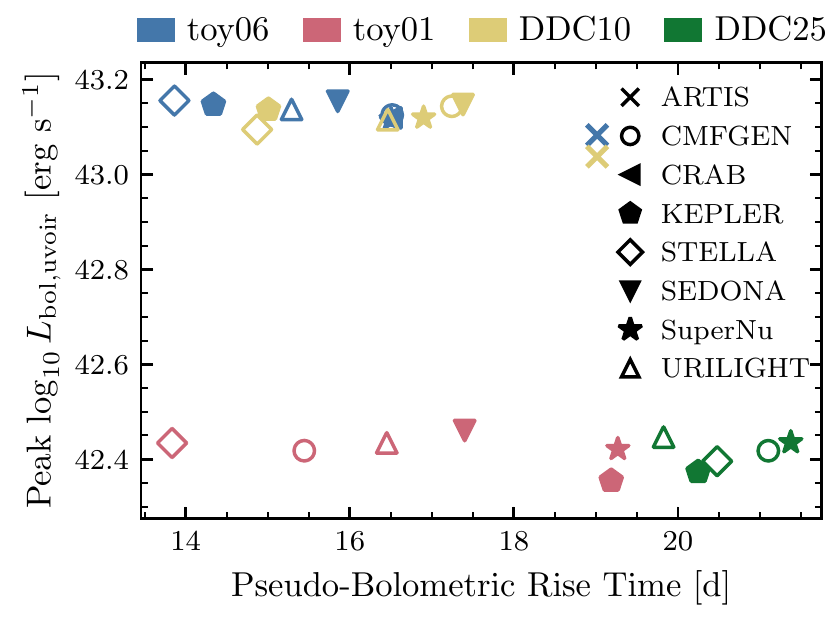}
 \caption{Peak pseudobolometric (UVOIR) luminosities vs. rise times
   (from explosion to peak) for all four test models. The data points
   corresponding to the CMFGEN, CRAB, and SuperNu calculations of the
   toy06 model are difficult to distinguish.
 }
 \label{fig:bolpeak}
\end{figure}

\begin{figure}
 \includegraphics[width=\hsize]{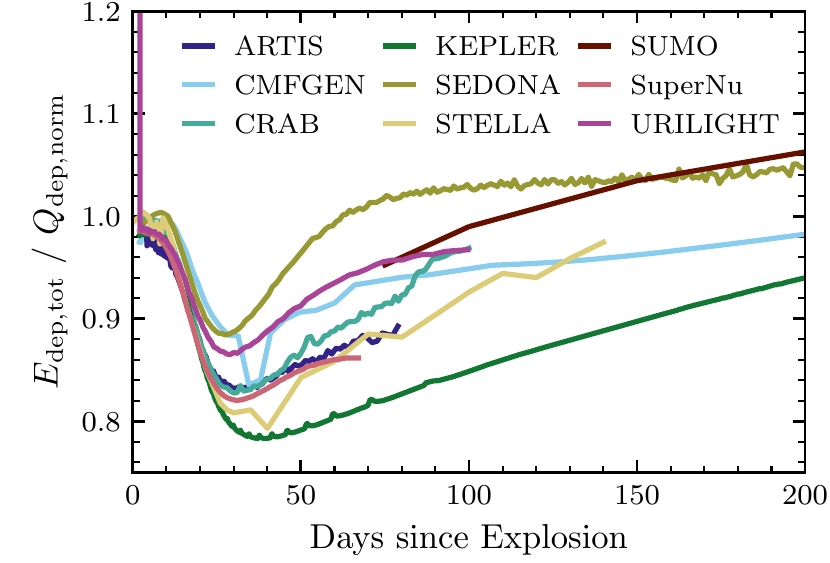}
   \caption{Total energy deposition rate from $\gamma$ rays and
     positrons for model toy06, normalised to the analytic function
     given by Eq. \ref{eq:Qdepnorm} (the normalisation allows
     differences to be seen more clearly).
   }
 \label{fig:edep}
\end{figure}

\begin{figure*}
 \centering
 \includegraphics[width=0.425\hsize]{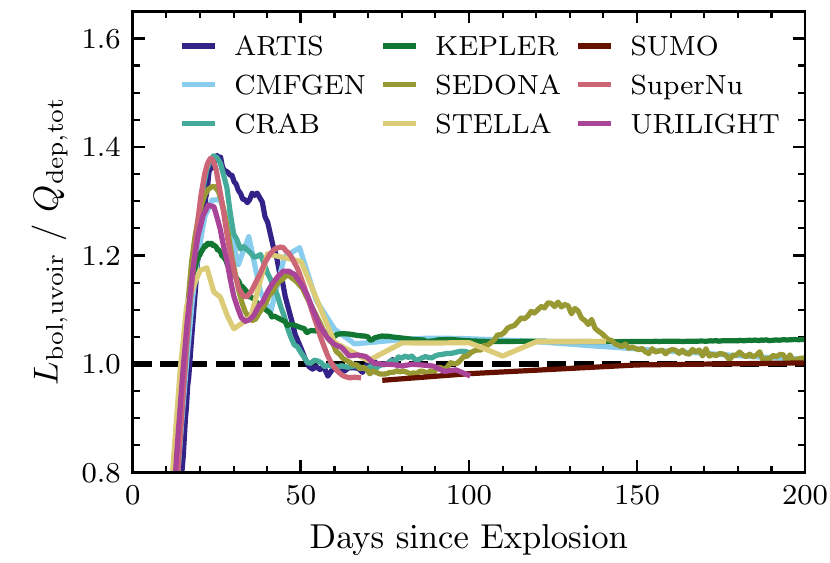}\hspace{.5cm}
 \includegraphics[width=0.425\hsize]{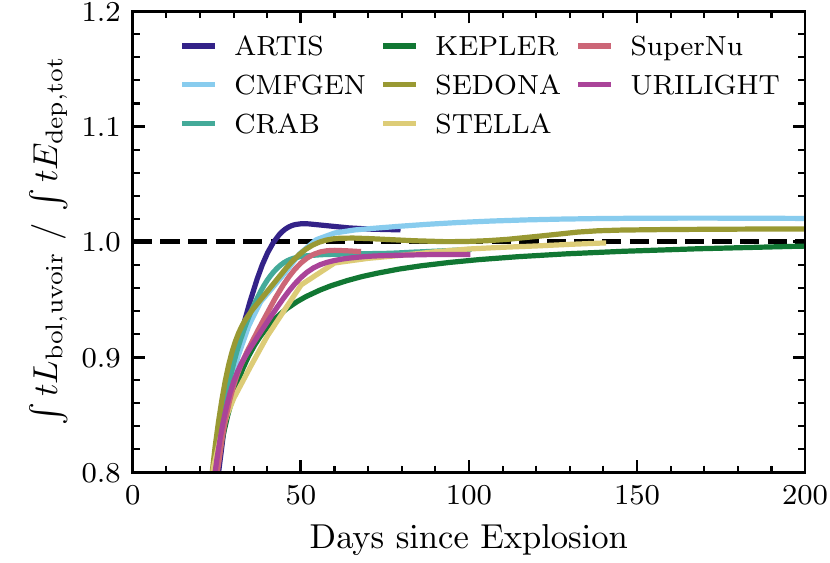}
 \caption{
 Left: Instantaneous ratio of the luminosity and the total energy
 deposition rate for model toy06.
 Right: Ratio of the time-weighted
 integrals of the luminosity and energy deposition rate for model
 toy06 (here we do not show the results for the SUMO code because the
 calculation starts at 75\,d post explosion).  Both ratios are
 expected to reach unity at late times.
}
 \label{fig:bolratios}
\end{figure*}

\begin{figure*}
 \centering
 \includegraphics[width=0.425\hsize]{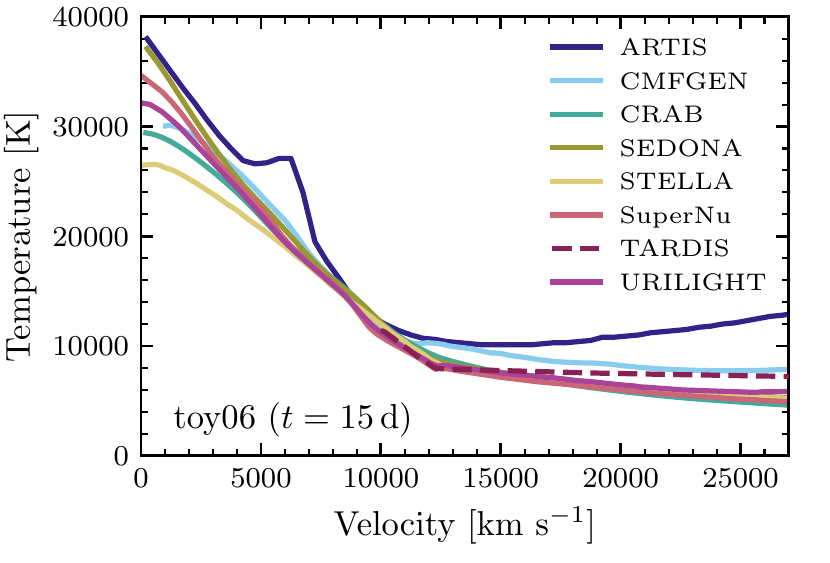}\hspace{.5cm}
 \includegraphics[width=0.425\hsize]{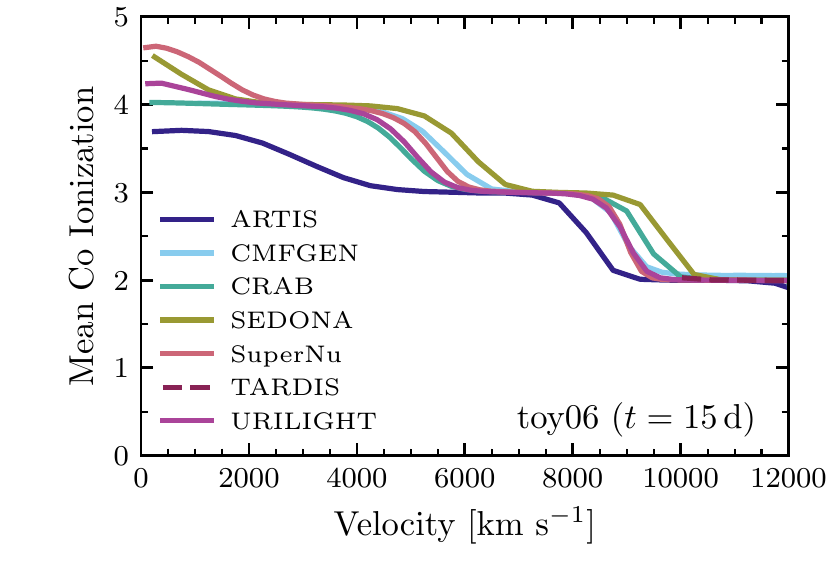}\vspace*{.1cm}
 \includegraphics[width=0.425\hsize]{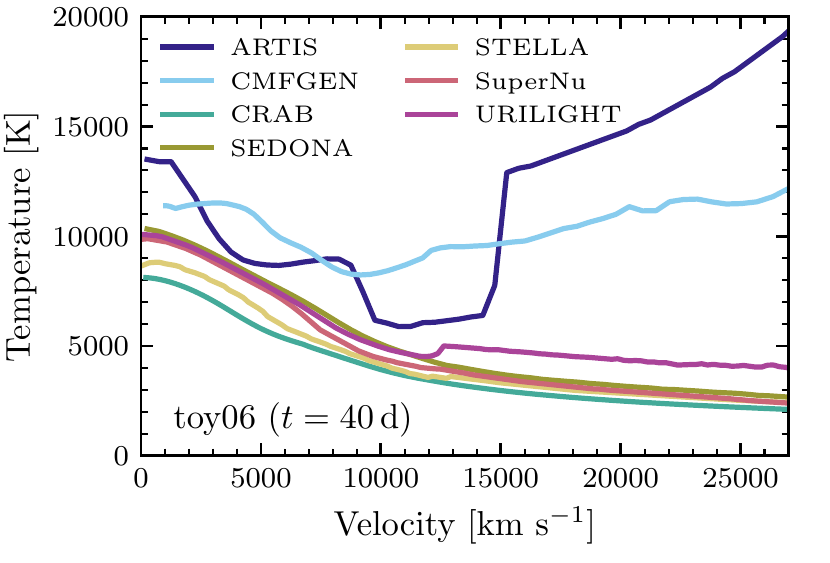}\hspace{.5cm}
 \includegraphics[width=0.425\hsize]{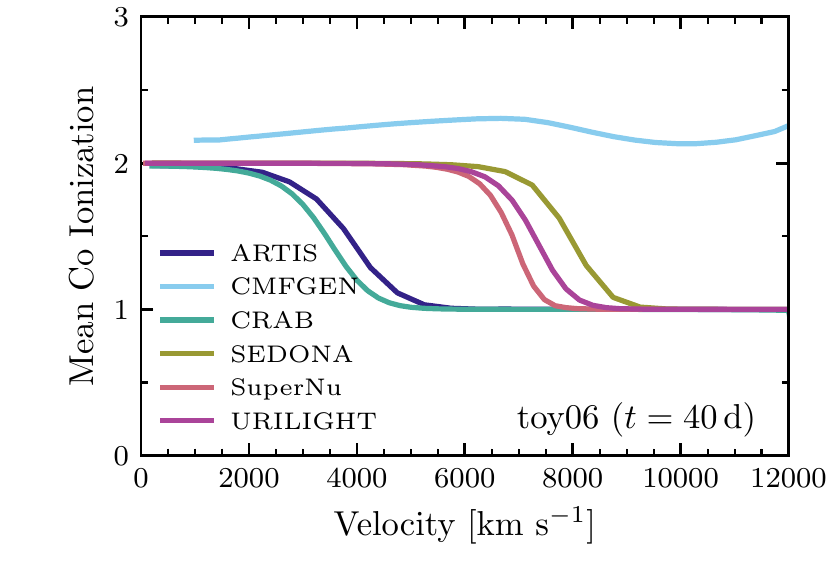}\vspace*{.1cm}
 \includegraphics[width=0.425\hsize]{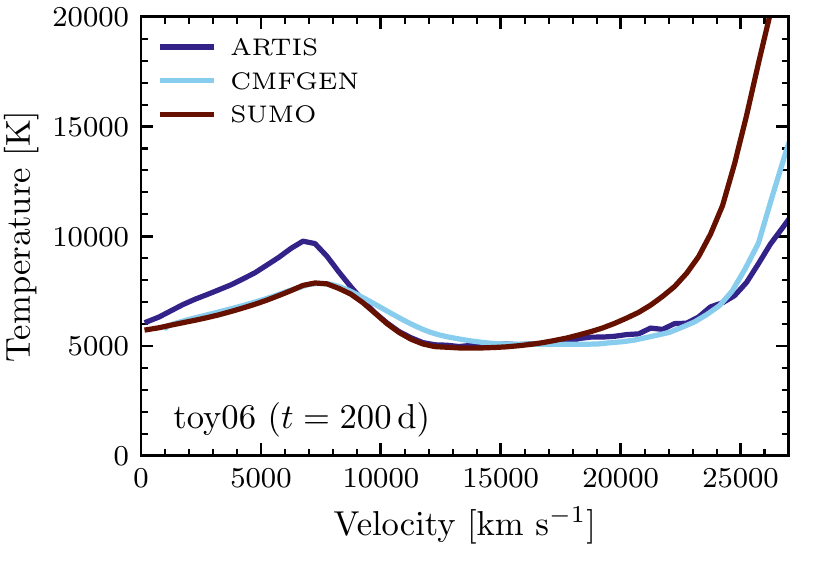}\hspace{.5cm} 
 \includegraphics[width=0.425\hsize]{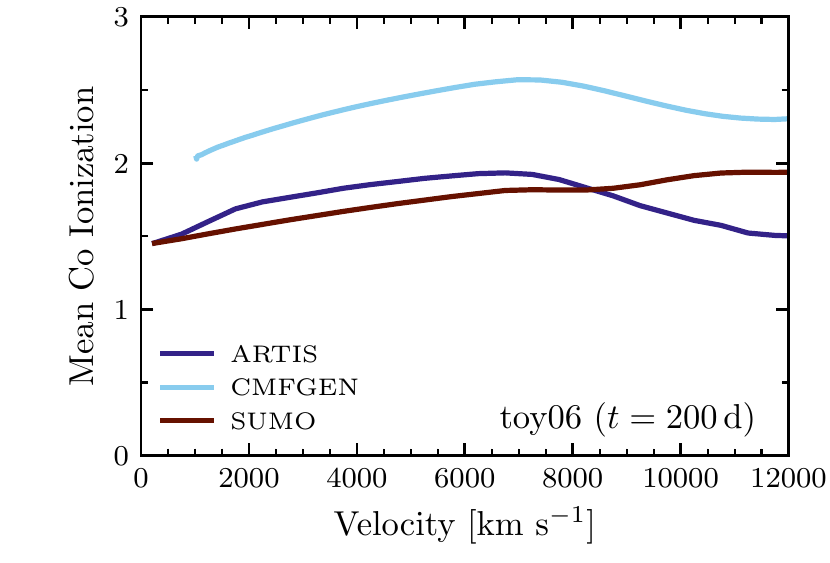}
 \caption{Ejecta (gas) temperature and ionisation profiles as a
   function of velocity for the toy06 model. Left column: Temperature
   at 15 days (top), 40 days (middle), and 200 days (bottom) post explosion. Right column: Mean ionisation of cobalt at 15 days
   (top), 40 days (middle), and 200 days (bottom) post explosion. We
   restrict the abscissa range of the ionisation plots to $v \le
   12000$\,\kms\ since the Co abundance drops to 0 at larger
   velocities.
 }
 \label{fig:temp_and_ionization}
\end{figure*}

\begin{figure*}
 \centering
 \includegraphics[width=0.85\hsize]{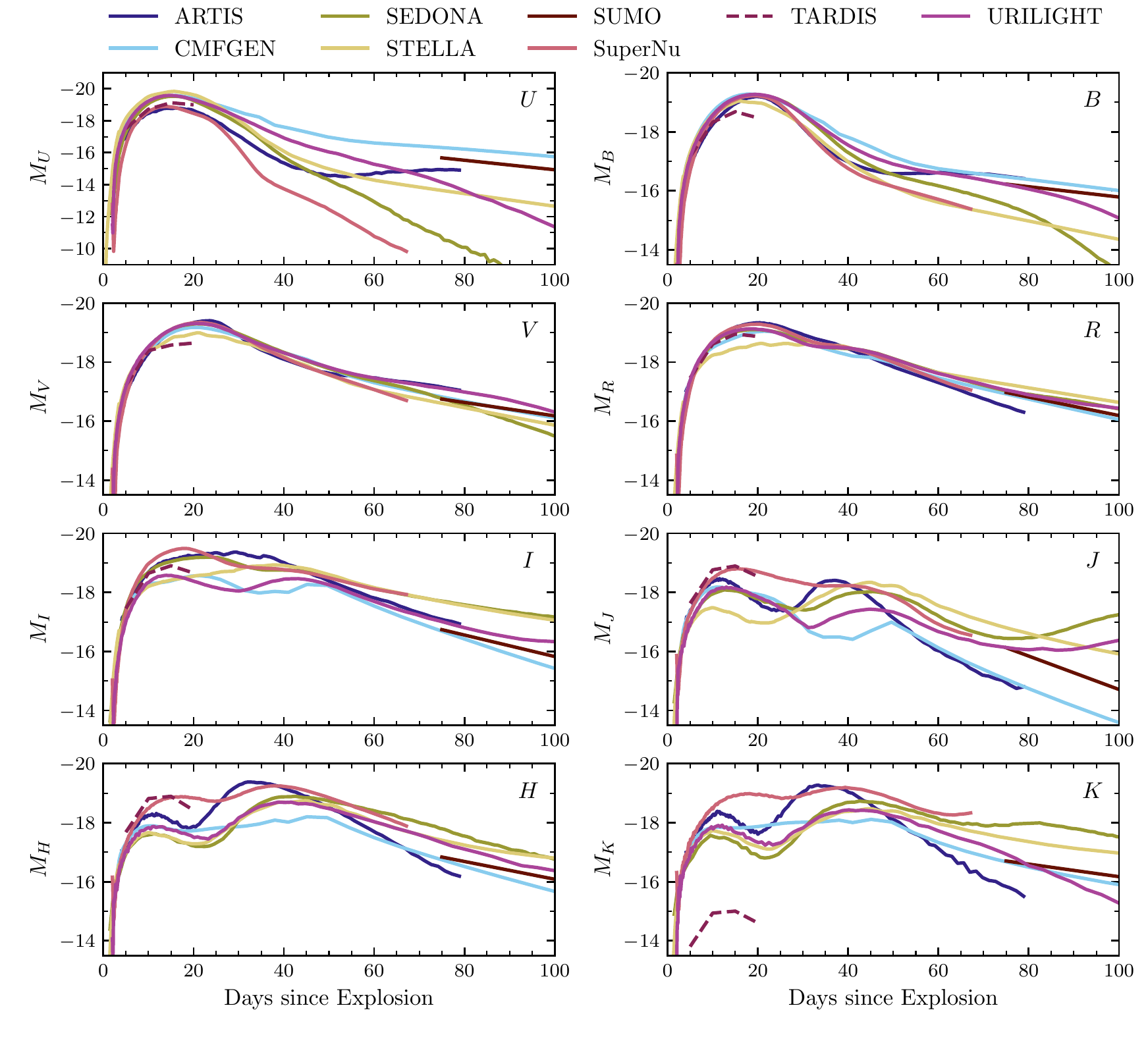}
 \caption{Multi-band ($UBVRIJHK$) light curves for the toy06
   model. We note the different ordinate range for the $U$ band.
 }
 \label{fig:lcfilt}
\end{figure*}

In this section we provide example results that are extracted from the
outputs of the different simulations. The purpose of this section is
to illustrate the contents of the data repository and comparisons that
can be made using it. No comparison to observations or in-depth
investigations of the sources of differences in results are made
here. We note that while some groups provided a few sets of results
for different physical approximations that appear in the repository,
only one set of results is shown for each code in this set of
examples. One exception is ARTIS, for which both the regular
(\texttt{artis}) and nebular (\texttt{artisnebular}) versions are
included for each model in order to present results for this code
from early to late times.

The results are provided in the following subsections. In
Sect. \ref{sect:bol}, bolometric properties of the emission are shown
for simulations of the toy06 model, including the pseudobolometric
(UVOIR) light curve and the energy deposition rate as a function of
time. Rise times and peak luminosities are provided for the toy06 and
toy01 models. In Sect. \ref{sect:tion}, the resulting profiles of
temperature and mean ionisation state of cobalt are shown at selected
times. In Sect. \ref{sect:multiband}, multi-band light curves and
colour curves are shown for model toy06. $B$-band rise times are
provided for all models where available. Example spectra at selected
early and late times are shown in Sect. \ref{sect:spec}.  Throughout
the section, each figure provides a legend with the colours or symbols
associated with each code. A uniform coding scheme is used, such that
a given code can be systematically identified in all the figures.

%%%%%%%%%%%%%%%%%%%%%%%%%%%%%%%%%%%%%%%%%%%%%%%%%%%%%%%%%%%%
\subsection{Bolometric (UVOIR) evolution}\label{sect:bol}

In this subsection, bolometric properties are shown based on the
\verb+lbol_edep+ output files. In Fig.~\ref{fig:lbol}, the UVOIR light
curves for model toy06 are shown for the first 100 days after
explosion for the different codes. The rise times and peak
luminosities are shown in Fig.~\ref{fig:bolpeak} for all four test
models. The peak luminosities were obtained by a parabolic fit to
light curve around maximum light.

In Fig.~\ref{fig:edep}, the total energy deposition from $\gamma$ rays
and positrons based on the same files is shown as a function of time
for the first 200 days after explosion for the toy06 model. In order
to highlight the differences among codes, all results are normalised
to the same analytic approximation for the deposition:

\begin{align}\label{eq:Qdepnorm}
Q_{\rm dep, norm}\equiv&\frac{M_{\rm Ni}}{M_{\odot}}\left(0.97\left[1 - e^{-(40d/t)^2}\right] + 0.03\right)\times\cr
&\left(6.45~e^{-t/8.8d}~+~1.45~e^{-t/111.3d}~\right)~10^{43}~\rm erg~s^{-1}
\end{align}

\noindent
where $M_{\rm Ni} = 0.6$\,\msun\ is the total (undecayed) $^{56}$Ni
mass in the toy06 model\footnote{The prefactor in parenthesis in the
first line of Eq. \eqref{eq:Qdepnorm}, is equivalent to a simplistic
approximation in which a fraction of 0.03 of the energy is emitted in
positron kinetic energy which is immediately deposited in the ejecta
while the rest is emitted in gamma-rays with a purely absorptive
optical depth given by $\tau_{\gamma}=t_0^2/t^2$ with a gamma-ray
escape time \citep[e.g.][]{Jefferey1999} of $t_0=40$d.}.

In Fig.~\ref{fig:bolratios}, two physical diagnostics of the relation
between the energy deposition and the bolometric light curves are
shown for the toy06 model (both expected to approach unity at late
times): the instantaneous ratio of the energy deposition rate and the
luminosity (left) and the ratio of cumulative time-weighted integrals
of the energy deposition rate and luminosity (right).

%%%%%%%%%%%%%%%%%%%%%%%%%%%%%%%%%%%%%%%%%%%%%%%%%%%%%%%%%%%%
\subsection{Temperature and ionisation profiles}\label{sect:tion}

In Fig.~\ref{fig:temp_and_ionization} the thermodynamic structure of
the ejecta as a function of velocity is shown based on the \verb+phys+
and \verb+ionfrac_co+ output files for model toy06. Three different
times of particular interest are shown: 15 days post explosion, close
to the peak of the light curve (upper panels), 40 days post explosion,
close to the break in the $B-V$ colour curve (middle panels), and 200
days post explosion, during the nebular phase (lower panels). For each
of these times, two profiles are shown: the (gas) temperature profile (left
panels), and the mean ionisation level of Co, defined as
$\sum\nolimits_{i=0} i\cdot f_i$, where $f_i$ is the fraction of Co
ions ionised $i$ times (right panels). Co and Fe dominate the opacity
at these times and the ionisation profiles of Fe are very similar to
those of Co (not shown here).

%%%%%%%%%%%%%%%%%%%%%%%%%%%%%%%%%%%%%%%%%%%%%%%%%%%%%%%%%%%%
\subsection{Multi-band light curves and colours}\label{sect:multiband}

\begin{figure}
 \centering
 \includegraphics[width=0.85\hsize]{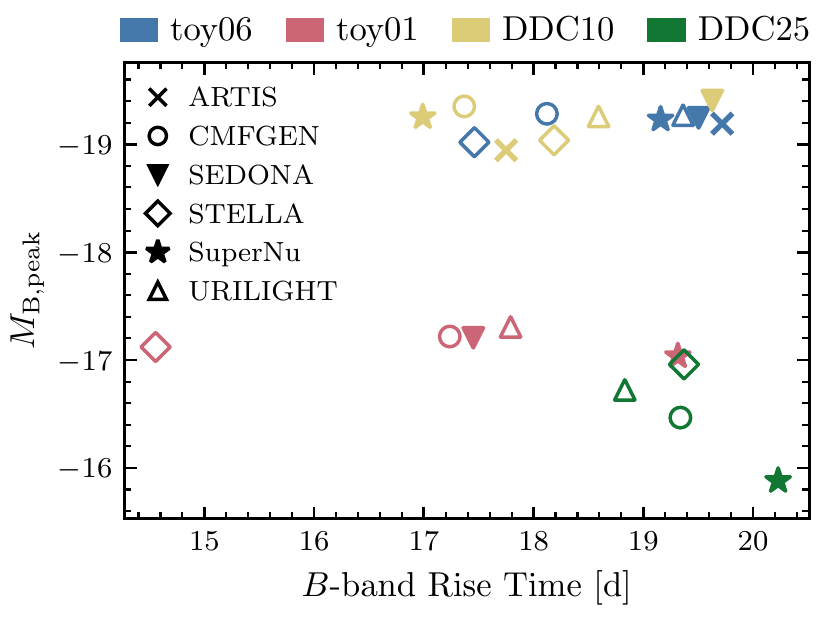}
 \caption{Peak magnitude vs. rise time in the $B$ band for all four test models.
}
 \label{fig:trise}
\end{figure}

\begin{figure*}
 \centering
 \includegraphics[width=0.425\hsize]{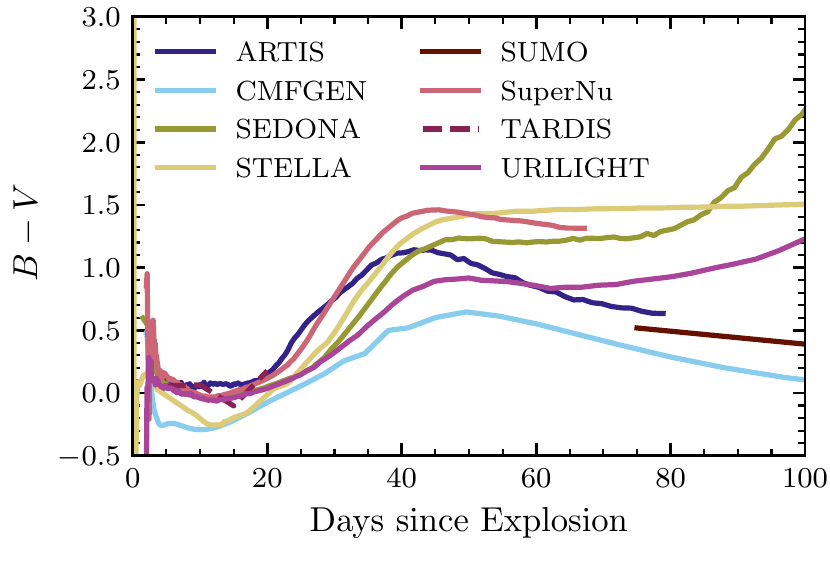}
 \includegraphics[width=0.425\hsize]{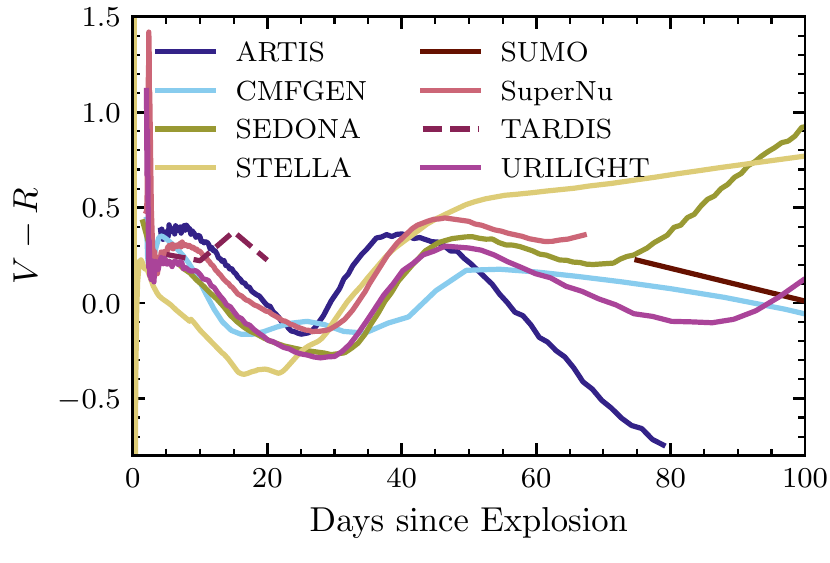}
 \caption{$B-V$ and $V-R$ colour evolution for the toy06 model.
 }
 \label{fig:color}
\end{figure*}

In this section, multi-band properties are shown for the different
codes for different models. The photometry is extracted from the
spectra reported in the \verb+spectra+ output files using the
\texttt{wsynphot}
package\footnote{\url{https://github.com/starkit/wsynphot}} (using the
Vega calibration spectrum
\verb+alpha_lyr_stis_003.fits+\footnote{\url{https://archive.stsci.edu/hlsps/reference-atlases/cdbs/calspec/alpha_lyr_stis_003.fits}}
and a third-order spline interpolation). In Fig.~\ref{fig:lcfilt}, the
$UBVRIJHK$ light curves are shown for model toy06 up to 100 days after
explosion. The $B$-band rise times (from explosion to peak) are
extracted by fitting a high-order polynomial around maximum light and
shown for all models in Fig.~\ref{fig:trise}. The $B-V$ and $V-R$
colour curves for model toy06 are shown in Fig.~\ref{fig:color}.

%%%%%%%%%%%%%%%%%%%%%%%%%%%%%%%%%%%%%%%%%%%%%%%%%%%%%%%%%%%%
\subsection{Spectroscopic evolution}\label{sect:spec}

% figures placed here to approach published version

\begin{figure*}
 \centering
 \includegraphics[width=0.425\hsize]{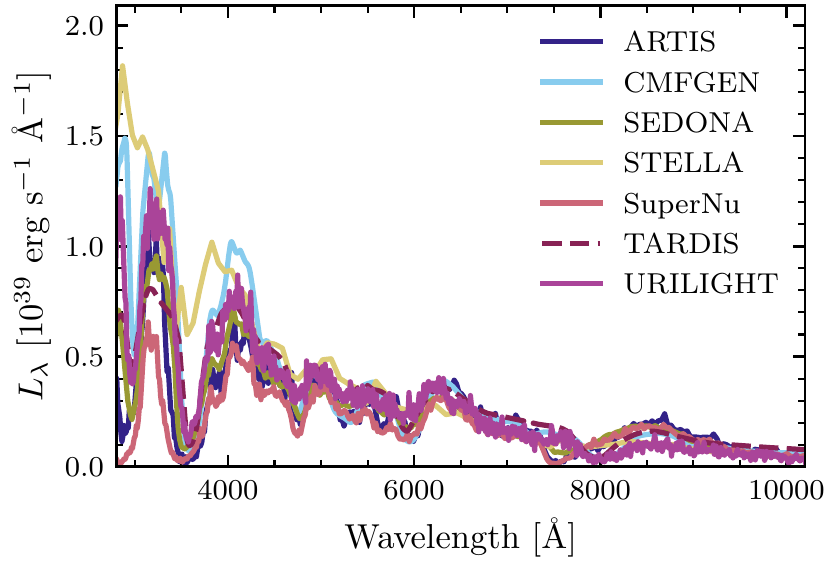}
 \includegraphics[width=0.425\hsize]{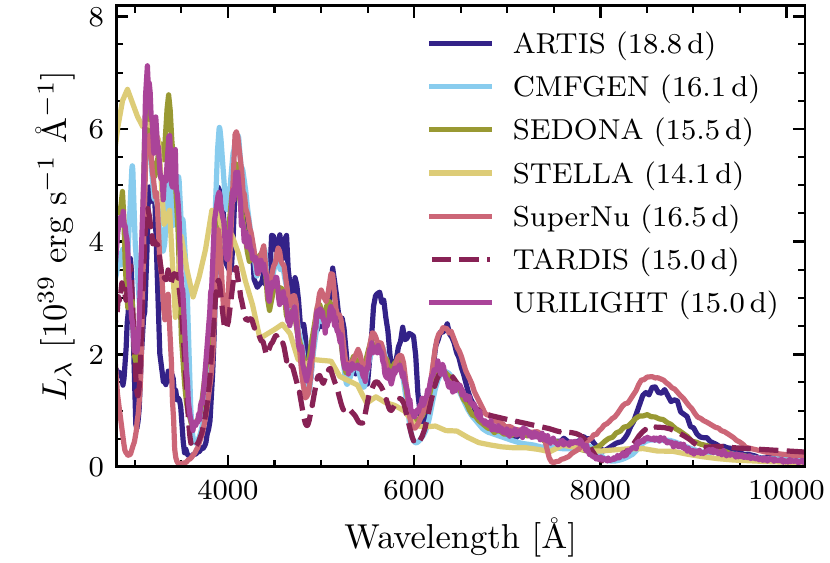}
 \caption{Spectra at 5\,d post explosion and at the time of peak UVOIR
   luminosity (the caption indicates the time of the spectrum computed
   closest to peak) for the toy06 model.
}
 \label{fig:earlyspec}
\end{figure*}

\begin{figure*}
 \centering
 \includegraphics[width=0.425\hsize]{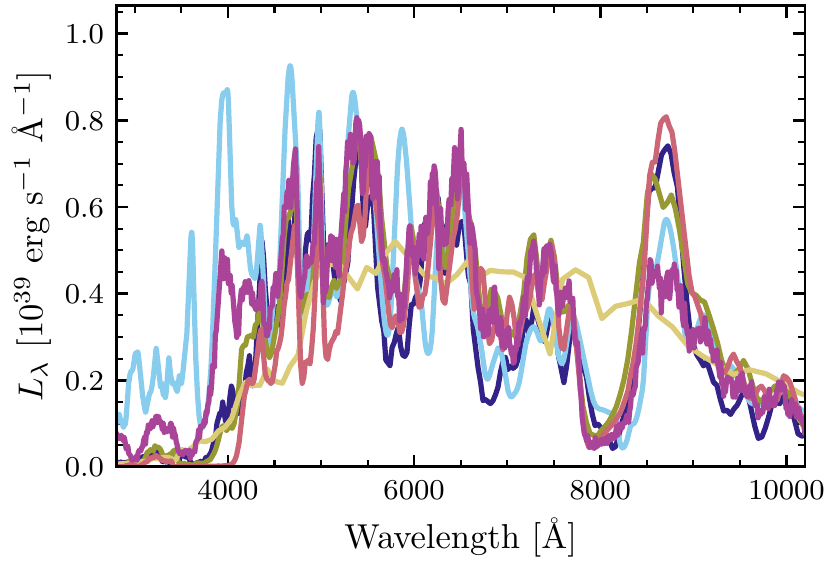}
 \includegraphics[width=0.425\hsize]{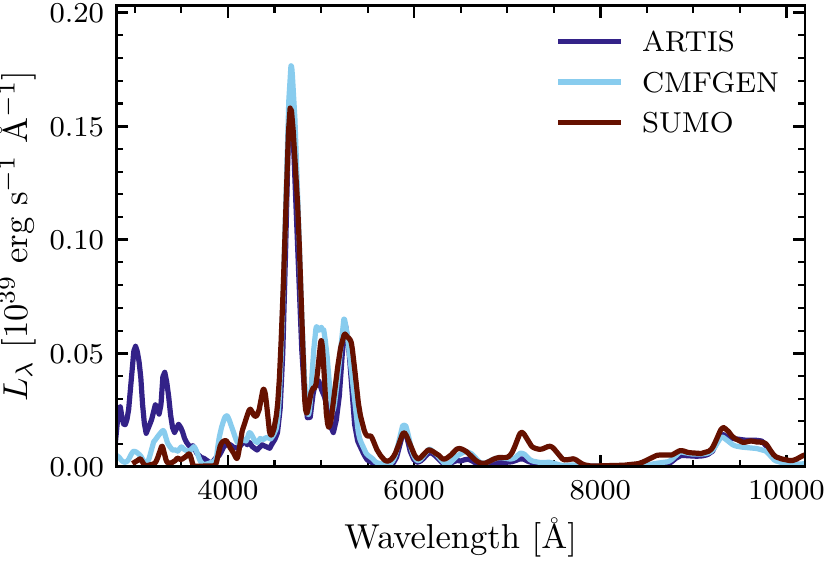}
 \caption{Spectra at 50\,d and 200\,d post explosion for the toy06
   model. The colour coding for the 50\,d spectra is the same as in
   Fig.~\ref{fig:earlyspec} and is not repeated here for sake of
   clarity.
}
 \label{fig:latespec}
\end{figure*}

Spectra obtained for the toy06 model at selected times are shown in
Figs.~\ref{fig:earlyspec} and \ref{fig:latespec} based on the
\verb+spectra+ output files. The wavelength range is restricted to
3000-10000\,\AA\ to allow direct comparison. This does not represent
the full range available in the output files, which differs between
different codes. In Fig.~\ref{fig:earlyspec}, early-time spectra (5
days post explosion and around peak luminosity) are shown while in
Fig.~\ref{fig:latespec} we show spectra at later times (50 days and
200 days post explosion). We note that nebular times require specialised
treatment in non-LTE which is only implemented in the ARTIS, CMFGEN,
and SUMO codes for this paper\footnote{SEDONA also has non-LTE
  capabilities but these were not used here.}.

%%%%%%%%%%%%%%%%%%%%%%%%%%%%%%%%%%%%%%%
% Discussion and Outlook
%%%%%%%%%%%%%%%%%%%%%%%%%%%%%%%%%%%%%%%

\section{Discussion and outlook}\label{sect:conclusions}

In this paper, we present an online public data repository that
includes test models for radiative-transfer (RT) calculations of
emission from SNe and standardised simulation outputs by ten
different groups that allows direct comparison. Python scripts that
generate the analytic toy models and read the different formats are
provided as well.

The main purpose of the repository is to allow studies of the
different physical approximations involved in the different codes and
to assess the robustness of different predictions of radiation
transfer. In addition, the repository can be used for finding bugs in
the codes or can provide checks in the development of new codes.

We plan to extend the set of test models to include other SN
types, as well as multi-dimensional models. In addition, we intend to
produce more specialised test cases for which exact solutions can be
found and agreed upon; these will provide benchmarks for RT code
development. We also aim to include models with standardised atomic
data sets that will allow the study of the effects of atomic physics
on the emission for the different approximations.

Other groups that are developing RT codes are encouraged to add their
results to the repository in the standardised format. Information for
this purpose is provided in the \texttt{README.md} file in the
repository.

%%%%%%%%%%%%%%%%%%%%%%%%%%%%%%%%%%%%%%%
% Acknowledgements
%%%%%%%%%%%%%%%%%%%%%%%%%%%%%%%%%%%%%%%

\begin{acknowledgements}
The results presented in this paper are based on work performed before Feb 24\textsuperscript{th} 2022.
We thank the Schwartz/Reisman Institute for Theoretical Physics at the Weizmann institute of science and the Max Plank Institute for Astrophysics for hosting the workshops during 2018-2019 which lead to this collaboration. 
This work was supported by the `Programme National de Physique Stellaire' (PNPS) of CNRS/INSU co-funded by CEA and CNES.
This research was supported by the Excellence Cluster ORIGINS which is funded by the Deutsche Forschungsgemeinschaft (DFG, German Research Foundation) under Germany's Excellence Strategy EXC-2094-390783311.
SB acknowledges support from the ESO Scientific Visitor Programme in Garching.
CC acknowledges support by the European Research Council (ERC) under the European Union's Horizon 2020 research and innovation programme under grant agreement No. 759253.
AJ acknowledges funding
from the European Research Council (ERC) under the European Union’s
Horizon 2020 Research and Innovation Program, ERC Starting Grant
803189 – SUPERSPEC.
Lawrence Livermore National Laboratory is operated by Lawrence Livermore National Security, LLC, for the U.S. Department of Energy, National Nuclear Security Administration under Contract DE-AC52-07NA27344.
AF and LS acknowledge support by the European Research Council (ERC) under the European Union’s Horizon 2020 research and innovation program (ERC Advanced Grant KILONOVA No. 885281).
The reported study was funded by the Russian Scientific Foundation (RSF),
project number 19-12-00229, and the Russian Foundation for Basic Research
(RFBR) and the Deutsche Forschungsgemeinschaft (DFG), project number 21-52-12032.
SAS acknowledges funding from the UKRI STFC Grant ST/T000198/1. Part of this work was performed using the Cambridge Service for Data Driven Discovery (CSD3), part of which is operated by the University of Cambridge Research Computing on behalf of the STFC DiRAC HPC Facility (www.dirac.ac.uk). The DiRAC component of CSD3 was funded by BEIS capital funding via STFC capital grants ST/P002307/1 and ST/R002452/1 and STFC operations grant ST/R00689X/1. DiRAC is part of the National e-Infrastructure.
M.W. acknowledges support from the NASA Future Investigators in NASA Earth and Space Science and Technology grant (80NSSC21K1849) and support from the Thomas J. Moore Fellowship at New York University.
Los Alamos National Laboratory is operated by Triad National Security, LLC, for the National Nuclear Security Administration of U.S. Department of Energy (Contract No. 89233218CNA000001). RTW also acknowledges Daniel van Rossum for the development of SuperNu that enabled this work.

\end{acknowledgements}

%%%%%%%%%%%%%%%%%%%% REFERENCES %%%%%%%%%%%%%%%%%%

\bibliographystyle{aa} % style aa.bst
\bibliography{main} % your references Yourfile.bib

%%%%%%%%%%%%%%%%%%%%%%%%%%%%%%%%%%%%%%%%%%%%%%%%%%%%%%%%%%%%%%%%%%%%%%

\begin{appendix}

%%%%%%%%%%%%%%%%%%%%%%%%%%%%%%%%%%%%%%%
% Atomic Data
%%%%%%%%%%%%%%%%%%%%%%%%%%%%%%%%%%%%%%%

\section{Atomic data}\label{sect:atomic_data}

In this Appendix we provide a more detailed overview of the atomic
data used in the calculations done with the ARTIS
(Sect.~\ref{sect:atomdata_artis}), CMFGEN
(Sect.~\ref{sect:atomdata_cmfgen}), SEDONA
(Sect.~\ref{sect:atomdata_sedona}), and TARDIS
(Sect.~\ref{sect:atomdata_tardis}) codes, along with appropriate
references. We refer the reader to Sect.~\ref{sect:codes} for a more
succinct presentation of the atomic data used by other groups and codes.

\subsection{ARTIS}\label{sect:atomdata_artis}

Table~\ref{tab:artisdata} lists the elements with numbers of ions,
levels, and bound-bound transitions used by the ARTIS code. These are
drawn from the Kurucz atomic line lists \citep[see ][]{Kromer2009}.

\begin{table}
\centering
\caption{Summary of atomic data used in ARTIS simulations.}\label{tab:artisdata}
\begin{tabular}{llrr} \hline
Element               &  Ion stages & Levels & Lines\\ \hline
{He} & {I-III}   & 723  & 1\,487\\
{C} & {I-V}      & 1\,032 & 7\,925\\
{N} & {I-V}      &  655 & 8\,991\\
{O} & {I-III}    & 381& 3\,388\\
{F} & {I-V}      & 752 & 7\,603\\
{Ne} & {I-V}     & 740 & 10\,462\\
{Na} & {I-V}     & 209 & 1\,249\\
{Mg} & {I-V}     & 724 & 4\,141\\
{Al} & {I-V}     & 556 & 4\,147\\
{Si} & {I-V}     & 1\,207 & 8\,984\\
{P} & {I-V}      & 436 & 2\,639\\
{S} & {I-V}      & 324 & 1\,805\\
{Cl} & {I-V}     & 469 & 5\,739\\
{Ar} & {I-V}     & 665 & 11\,629\\
{K} & {I-V}      & 167 & 1\,031\\
{Ca} & {I-V}     & 613 & 13\,003\\
{Sc} & {I-VII}   & 692 & 11\,863\\
{Ti} & {I-VII}   & 993 & 22\,139\\
{V} & {I-VII}    & 1\,343 & 32\,273\\
{Cr} & {I-VII}   & 1\,777 & 53\,557\\
{Mn} & {I-VII}   & 1\,726 & 38\,041\\
{Fe} & {I-VII}   & 2\,760 & 111\,890\\
{Co} & {I-VII}   & 1\,529 & 34\,098\\
{Ni} & {I-VII}   & 2\,078 & 54\,990\\
{Cu} & {I-V}     & 360 & 5\,072\\
{Zn} & {I-V}    & 313 & 3\,178\\ 
\textbf{Total} &  & \textbf{23\,224} & \textbf{461\,324} \\
\hline
\end{tabular}
\end{table}

\subsection{CMFGEN}\label{sect:atomdata_cmfgen}

Tables~\ref{tab:cmfgen_atoms_toy06}-\ref{tab:cmfgen_atoms_ddc_reducednicofe4} give the number of levels (both super-levels and full levels; see
\citealt{Hillier1998_cmfgen} and \citealt{Dessart2010} for details)
for the model atoms used in the radiative-transfer calculations
presented in this paper.  $N_\mathrm{SL}$ refers to the number of
super levels used for the solution of the rate equations, and
$N_\mathrm{full}$ refer the number of full levels used to solve the
transfer equation and compute the observed spectrum.  We report the
uppermost level for each ion treated in the fourth column.  `W' refers
to states in which higher $\ell$ states (usually f or higher) have
been combined into a single level.  In the last column we give the
number of bound-bound transitions in the model ion taking into account
all $N_\mathrm{full}$ levels.  Ions for which $N_\mathrm{SL} =
N_\mathrm{full} = 1$ (with no entries for the last level configuration
or number of lines) correspond to the final ionisation stage of a
given element, for which ionisations to and recombinations from the
ground state are considered.  The ions Cl\four, K\three, and
V\one\ were included for the sole purpose of tracking changes in
abundance of radioactive isotopes. The entries for those ions in
Tables~\ref{tab:cmfgen_atoms_toy06}-\ref{tab:cmfgen_atoms_ddc_early}
also have $N_\mathrm{SL} = N_\mathrm{full} = 1$, but we specify the
configuration of the ground state in the last level column and set the
number of lines to zero.

Oscillator strengths for CNO elements were originally taken from
\cite{Nussbaumer1983_LTDR, Nussbaumer1984_CNO_LTDR}. These authors
also provide transition probabilities to states in the ion
continuum. The largest source of oscillator data is from
\cite{Kurucz2009_ATD}\footnote{Data are available online at
\url{http://kurucz.harvard.edu}}; its principal advantage over many
other sources (e.g. Opacity Project) is that LS coupling is not
assumed. More recently, non-LS oscillator strengths have become
available through the Iron Project \citep{Hummer1993_IP}, and work
done by the atomic-data group at Ohio State University
\citep{Nahar2010_OSU}. Other important sources of radiative data for
Fe include \cite{Becker1992_FeV, Becker1995_FeVI, Becker1995_FeIV,
  Nahar1995_FeII}.  Energy levels have generally been obtained from
the National Institute of Standards and Technology. Collisional data
are sparse, particularly for states far from the ground state. The
principal source for collisional data among low-lying states for a
variety of species is the tabulation by \cite{Mendoza1983_col}; other
sources include \cite{Berrington1985_col}, \cite{Lennon1985_CII_col},
\cite{Lennon1994_N2}, \cite{Shine1974},
\cite{Tayal1997_SII_col,Tayal1997_SIII_col},
\cite{Zhang1995_FeII_col,Zhang1995_FeIII_col,Zhang1997_FeIV_col}.
Photoionisation data is taken from the Opacity Project
\citep{Seaton1987_OP} and the Iron Project
\citep{Hummer1993_IP}. Unfortunately Ni and Co photoionisation data are
generally unavailable, and we used crude
approximations. Charge exchange cross-sections are from the tabulation
by \cite{Kingdon1996_chg}.

\begin{table}
\centering
\caption{Model atoms used in CMFGEN calculations for the toy06 model.} % based on snia_toy06_17
\label{tab:cmfgen_atoms_toy06}
\begin{tabular}{lrrlr}
\hline
Ion & $N_\mathrm{SL}$ & $N_\mathrm{full}$ & Last level & $N_\mathrm{lines}$  \\
\hline
\ion{Si}{2}    &        32  &          62  &   3s$^2$ 7g $^2$G                                   &          1\,032     \\ % 3s2_7g_2Ge[7/2]
\ion{Si}{3}    &        33  &          61  &   3s 5g$^3$G                                        &             615     \\ % 3s5g3Ge[5]
\ion{Si}{4}    &        37  &          48  &   10f $^2$F$^{\sr o}$                               &             816     \\ % 10f_2Fo
\ion{Si}{5}    &         1  &           1  &                                                     &                     \\
\ion{S}{2}     &        56  &         324  &   3s 3p$^3$($^5$S$^{\sr o}$)4p $^6$P                &         16\,965     \\ % 3s_3p3(5So)4p_6Pe
\ion{S}{3}     &        48  &          98  &   3s 3p$^2$($^2$D)3d $^3$P                          &          1\,723     \\ % 3s_3p2(2De)3d_3Pe
\ion{S}{4}     &        27  &          67  &   3s 3p($^3$P$^{\sr o}$)4p $^2$D                    &          1\,091     \\ % 3s_3p(3Po)4p_2De[5/2]
\ion{S}{5}     &         1  &           1  &                                                     &                     \\
\ion{Ca}{2}    &        44  &          77  &   3p$^6$ 30w $^2$W                                  &          3\,365     \\ % 3p6_30w_2W
\ion{Ca}{3}    &        16  &          40  &   3s$^2$ 3p$^5$ 5s $^1$P$^{\sr o}$                  &             210     \\ % 3s2_3p5_5s_1Po
\ion{Ca}{4}    &        18  &          69  &   3s 3p$^5$($^3$P$^{\sr o}$)3d $^4$D$^{\sr o}$      &             646     \\ % 3s_3p5(3Po)3d_4Do[1/2]
\ion{Ca}{5}    &         1  &           1  &                                                     &                     \\
\ion{Fe}{1}    &        44  &         136  &   3d6($^5$D)4s 4p x$^5$F$^{\sr o}$                  &          3\,934     \\ % 3d6(5D)4s4p_x5Fo[3]
\ion{Fe}{2}    &       228  &      2\,698  &   3d$^5$($^4$F)4s 4p b$^4$G$^{\sr o}$               &     1\,060\,528     \\ % 3d5(4F)4s_4p_b4Go[7/2]
\ion{Fe}{3}    &        83  &         698  &   3d$^5$($^2$H)4d 1K                                &         73\,419     \\ % 3d5(2H)4d_1Ke[7]
\ion{Fe}{4}    &       100  &      1\,000  &   3d$^4$($^3$G)4f $^4$P$^{\sr o}$                   &        144\,005     \\ % 3d4(3G)4f_4Po[5/2]
\ion{Fe}{5}    &        47  &         191  &   3d$^3$($^4$F)4d $^5$F                             &          7\,892     \\ % 3d3(4F)4d_5Fe[3]
\ion{Fe}{6}    &         1  &           1  &                                                     &                     \\
\ion{Co}{2}    &       136  &      2\,747  &   3d$^7$($^2$D)6p $^3$P$^{\sr o}$                   &     1\,186\,076     \\ % 3d7(2D)6p_3Po[1]
\ion{Co}{3}    &       124  &      3\,917  &   3d$^6$($^3$D)6d $^4$P                             &     1\,357\,405     \\ % 3d6(3D)6d_4Pe[3/2]
\ion{Co}{4}    &        37  &         314  &   3d$^5$($^2$P)4p $^3$P$^{\sr o}$                   &         17\,952     \\ % 3d5(2P)4p_3Po[1]
\ion{Co}{5}    &        32  &         387  &   3d$^4$($^3$F)4d $^2$H                             &         27\,046     \\ % 3d4(3F)4d_2He[9/2]
\ion{Co}{6}    &         1  &           1  &                                                     &                     \\
\ion{Ni}{2}    &        59  &      1\,000  &   3d$^8$($^3$F)7f $^4$I$^{\sr o}$                   &        103\,224     \\ % 3d8(3F)7f_4Io[9/2]
\ion{Ni}{3}    &        47  &      1\,000  &   3d$^7$($^2$D)4d $^3$Sb                            &        132\,677     \\ % 3d7(2D)4d_3Sbe[1]
\ion{Ni}{4}    &        28  &         254  &   3d$^6$($^1$G1)4p $^2$G$^{\sr o}$                  &         12\,512     \\ % 3d6(1G1)4p_2Go[7/2]
\ion{Ni}{5}    &        46  &         183  &   3d$^5$($^2$D3)4p $^3$F$^{\sr o}$                  &          6\,033     \\ % 3d5(2D3)4p_3Fo[3]
\ion{Ni}{6}    &         1  &           1  &                                                     &                     \\
\textbf{Total} & \textbf{1\,328} & \textbf{15\,377} &                                            & \textbf{4\,159\,166}\\
\hline
\end{tabular}
\flushleft
\textbf{Notes:} 
Due to a $gf$ cut (level dependent, $gf > 10^{-4}$) only 1\,292\,015
lines were included in the non-LTE calculations of the level
populations. 2\,082\,610 lines were included when computing the
observed spectrum. Prior to 4.25\,d post explosion a higher cut was
used ($gf > 10^{-3}$) in order to ease convergence. From 28.52\,d
onward we omitted the highest ionisation stages Fe\,\five, Co\,\five,
and Ni\,\five\ because their populations are too low to affect the
radiative transfer.
\end{table}

\begin{table}
\centering
\caption{Model atoms used in CMFGEN calculations for the toy01 model.} % based on snia_toy01_11
\label{tab:cmfgen_atoms_toy01}
\begin{tabular}{lrrlr}
\hline
Ion & $N_\mathrm{SL}$ & $N_\mathrm{full}$ & Last level & $N_\mathrm{lines}$  \\
\hline
\ion{Si}{2}    &        32  &          62  &   3s$^2$ 7g $^2$G                                   &          1\,032     \\ % 3s2_7g_2Ge[7/2]
\ion{Si}{3}    &        33  &          61  &   3s 5g$^3$G                                        &             615     \\ % 3s5g3Ge[5]
\ion{Si}{4}    &        37  &          48  &   10f $^2$F$^{\sr o}$                               &             816     \\ % 10f_2Fo
\ion{Si}{5}    &         1  &           1  &                                                     &                     \\
\ion{S}{2}     &        56  &         324  &   3s 3p$^3$($^5$S$^{\sr o}$)4p $^6$P                &         16\,965     \\ % 3s_3p3(5So)4p_6Pe
\ion{S}{3}     &        48  &          98  &   3s 3p$^2$($^2$D)3d $^3$P                          &          1\,723     \\ % 3s_3p2(2De)3d_3Pe
\ion{S}{4}     &        27  &          67  &   3s 3p($^3$P$^{\sr o}$)4p $^2$D                    &          1\,091     \\ % 3s_3p(3Po)4p_2De[5/2]
\ion{S}{5}     &         1  &           1  &                                                     &                     \\
\ion{Ca}{2}    &        44  &          77  &   3p$^6$ 30w $^2$W                                  &          3\,365     \\ % 3p6_30w_2W
\ion{Ca}{3}    &        16  &          40  &   3s$^2$ 3p$^5$ 5s $^1$P$^{\sr o}$                  &             210     \\ % 3s2_3p5_5s_1Po
\ion{Ca}{4}    &        18  &          69  &   3s 3p$^5$($^3$P$^{\sr o}$)3d $^4$D$^{\sr o}$      &             646     \\ % 3s_3p5(3Po)3d_4Do[1/2]
\ion{Ca}{5}    &         1  &           1  &                                                     &                     \\
\ion{Fe}{1}    &        44  &         136  &   3d6($^5$D)4s 4p x$^5$F$^{\sr o}$                  &          3\,934     \\ % 3d6(5D)4s4p_x5Fo[3]
\ion{Fe}{2}    &       228  &      2\,698  &   3d$^5$($^4$F)4s 4p b$^4$G$^{\sr o}$               &     1\,060\,528     \\ % 3d5(4F)4s_4p_b4Go[7/2]
\ion{Fe}{3}    &        83  &         698  &   3d$^5$($^2$H)4d 1K                                &         73\,419     \\ % 3d5(2H)4d_1Ke[7]
\ion{Fe}{4}    &       100  &      1\,000  &   3d$^4$($^3$G)4f $^4$P$^{\sr o}$                   &        144\,005     \\ % 3d4(3G)4f_4Po[5/2]
\ion{Fe}{5}    &        47  &         191  &   3d$^3$($^4$F)4d $^5$F                             &          7\,892     \\ % 3d3(4F)4d_5Fe[3]
\ion{Fe}{6}    &         1  &           1  &                                                     &                     \\
\ion{Co}{2}    &        44  &         162  &   3d$^6$($^5$D)4s 4p $^7$F$^{\sr o}$                &          6\,475     \\ % 3d6(5D)4s_4p_7Fo[0]
\ion{Co}{3}    &        33  &         220  &   3d$^6$($^3$F)4p $^4$D$^{\sr o}$                   &          9\,836     \\ % 3d6(3F)4p_4Do[7/2]
\ion{Co}{4}    &        27  &         164  &   3d$^5$($^4$D)4p $^3$F$^{\sr o}$                   &          5\,759     \\ % 3d5(4D)4p_3Fo[2]
\ion{Co}{5}    &        32  &         387  &   3d$^4$($^3$F)4d $^2$H                             &         27\,046     \\ % 3d4(3F)4d_2He[9/2]
\ion{Co}{6}    &         1  &           1  &                                                     &                     \\
\ion{Ni}{2}    &        27  &         177  &   3d7($^4$P)4s 4p($^3$P) $^4$S$^{\sr o}$            &          5\,757     \\ % 3d7(4P)4s4p(3P)_4So[3/2]
\ion{Ni}{3}    &        20  &         107  &   3d$^7$($^2$H)4p $^3$I$^{\sr o}$                   &          2\,228     \\ % 3d7(2H)4p_3Io[6]
\ion{Ni}{4}    &        20  &         130  &   3d$^6$($^3$F2)4p $^4$D$^{\sr o}$                  &          3\,375     \\ % 3d6(3F2)4p_4Do[7/2]
\ion{Ni}{5}    &        46  &         183  &   3d$^5$($^2$D3)4p $^3$F$^{\sr o}$                  &          6\,033     \\ % 3d5(2D3)4p_3Fo[3]
\ion{Ni}{6}    &         1  &           1  &                                                     &                     \\
\textbf{Total} & \textbf{1\,068} & \textbf{7\,105}  &                                            & \textbf{1\,382\,750}\\
\hline
\end{tabular}
\flushleft
\textbf{Notes:} 
Due to a $gf$ cut (level dependent, $gf > 10^{-4}$) only 377\,383
lines were included in the non-LTE calculations of the level
populations. 692\,911 lines were include when computing the observed
spectrum. Prior to 2.4\,d post explosion a higher cut was used ($gf >
10^{-3}$) in order to ease convergence. From 25.93\,d onward we
omitted the highest ionisation stages Fe\,\five, Co\,\five, and
Ni\,\five\ since their populations are too low to affect the radiative
transfer.
\end{table}

\clearpage

\xentrystretch{-.1}
\topcaption{Model atoms used in CMFGEN calculations up until 10.56\,d post explosion for the DDC10 and DDC25 models.}% based on DDC10_0p5_33
\label{tab:cmfgen_atoms_ddc_early}
\tablefirsthead{\toprule Ion & $N_\mathrm{SL}$ & $N_\mathrm{full}$ & Last level & $N_\mathrm{lines}$ \\ \midrule}
\tablehead{
\multicolumn{5}{l}{\footnotesize \textbf{Table~\ref*{tab:cmfgen_atoms_ddc_early}.} continued.} \\
\multicolumn{5}{l}{} \\
\toprule
Ion & $N_\mathrm{SL}$ & $N_\mathrm{full}$ & Last level & $N_\mathrm{lines}$ \\ 
\midrule
}
\tabletail{\midrule}
\tablelasttail{\bottomrule}
\begin{xtabular}{lrrlr}
\ion{C}{1}     &        14  &          26  &   2s 2p$^3$ $^3$P$^{\sr o}$                         &             229     \\ % 2s_2p3_3Po[0]
\ion{C}{2}     &        14  &          26  &   2s 2s4d$^2$D                                      &             159     \\ % 2s2s4d2De[5/2]
\ion{C}{3}     &        62  &         112  &   2s 8f$^1$F$^{\sr o}$                              &          1\,759     \\ % 2s8f1Fo
\ion{C}{4}     &        59  &          64  &   30                                                &          2\,798     \\ % 30___
\ion{C}{5}     &         1  &           1  &                                                     &                     \\
\ion{O}{1}     &        19  &          51  &   2s$^2$ 2p$^3$($^4$S$^{\sr o}$)4f $^3$F            &             401     \\ % 2s2_2p3(4So)4f_3Fe[3]
\ion{O}{2}     &        30  &         111  &   2s$^2$ 2p$^2$($^3$P)4d $^2$D                      &          2\,242     \\ % 2s2_2p2(3Pe)4d_2De[5/2]
\ion{O}{3}     &        50  &          86  &   2p 4f$^1$D                                        &          1\,270     \\ % 2p4f1De
\ion{O}{4}     &        53  &          72  &   2p 2p3p''$^2$P$^{\sr o}$                          &          1\,614     \\ % 2p2p3p''2Po
\ion{O}{5}     &         1  &           1  &                                                     &                     \\
\ion{Ne}{1}    &        70  &         139  &   2s$^2$ 2p$^5$($^2$P<3/2>)6d 2{5/2}$^{\sr o}$      &          3\,126     \\ % 2s2_2p5(2P<3/2>)6d_2{5/2}o[3]
\ion{Ne}{2}    &        22  &          91  &   2s$^2$ 2p$^4$($^3$P)4d $^2$P                      &          2\,142     \\ % 2s2_2p4(3Pe)4d_2Pe[3/2]
\ion{Ne}{3}    &        23  &          71  &   2s$^2$ 2p$^3$($^2$D$^{\sr o}$)3d $^3$S$^{\sr o}$  &             884     \\ % 2s2_2p3(2Do)3d_3So[1]
\ion{Ne}{4}    &         1  &           1  &                                                     &                     \\
\ion{Na}{1}    &        22  &          71  &   30w$^2$W                                          &          3\,128     \\ % 30w2W
\ion{Na}{2}    &         1  &           1  &                                                     &                     \\
\ion{Mg}{2}    &        22  &          65  &   30w$^2$W                                          &          2\,810     \\ % 30w2W
\ion{Mg}{3}    &        31  &          99  &   2p$^5$ 7s $^1$P$^{\sr o}$                         &          1\,526     \\ % 2p5_7s_1Po
\ion{Mg}{4}    &         1  &           1  &                                                     &                     \\
\ion{Al}{2}    &        26  &          44  &   3s 5d$^1$D                                        &             333     \\ % 3s5d1De[2]
\ion{Al}{3}    &        17  &          45  &   10z$^2$Z                                          &             699     \\ % 10z2Z
\ion{Al}{4}    &         1  &           1  &                                                     &                     \\
\ion{Si}{2}    &        31  &          59  &   3s$^2$($^1$S)7g$^2$G                              &             683     \\ % 3s2(1S)7g2Ge[7/2]
\ion{Si}{3}    &        33  &          61  &   3s 5g$^1$G                                        &             614     \\ % 3s5g1Ge[4]
\ion{Si}{4}    &        37  &          48  &   10f$^2$F$^{\sr o}$                                &             781     \\ % 10f2Fo
\ion{Si}{5}    &         1  &           1  &                                                     &                     \\
\ion{S}{2}     &        56  &         324  &   3s 3p$^3$($^5$S$^{\sr o}$)4p $^6$P                &         16\,346     \\ % 3s_3p3(5So)4p_6Pe
\ion{S}{3}     &        48  &          98  &   3s 3p$^2$($^2$D)3d $^3$P                          &          1\,629     \\ % 3s_3p2(2De)3d_3Pe
\ion{S}{4}     &        27  &          67  &   3s 3p($^3$P$^{\sr o}$)4p $^2$D                    &             760     \\ % 3s_3p(3Po)4p_2De[5/2]
\ion{S}{5}     &         1  &           1  &                                                     &                     \\
\ion{Cl}{4}    &         1  &           1  &   3p$^2$ $^3$P                                      &               0     \\ % 3p2_3Pe[0]
\ion{Cl}{5}    &         1  &           1  &                                                     &                     \\
\ion{Ar}{1}    &        56  &         110  &   3s$^2$ 3p$^5$($^2$P<3/2>)7p 2{3/2}                &          3\,030     \\ % 3s2_3p5(2P<3/2>)7p_2{3/2}e[2]
\ion{Ar}{2}    &       134  &         415  &   3s$^2$ 3p$^4$($^3$P<1>)7i 2{6}                    &         40\,224     \\ % 3s2_3p4(3P<1>)7i_2{6}e[11/2]
\ion{Ar}{3}    &        32  &         346  &   3s$^2$ 3p$^3$($^2$D$^{\sr o}$)8s $^1$D$^{\sr o}$  &         13\,677     \\ % 3s2_3p3(2Do)8s_1Do
\ion{Ar}{4}    &         1  &           1  &                                                     &                     \\
\ion{K}{3}     &         1  &           1  &   3p$^5$ $^2$P$^{\sr o}$                            &               0     \\ % 3p5_2Po[3/2]
\ion{K}{4}     &         1  &           1  &                                                     &                     \\
\ion{Ca}{2}    &        21  &          77  &   3p$^6$ 30w$^2$W                                   &          3\,365     \\ % 3p6_30w2W
\ion{Ca}{3}    &        16  &          40  &   3s$^2$ 3p$^5$ 5s $^1$P$^{\sr o}$                  &             210     \\ % 3s2_3p5_5s_1Po
\ion{Ca}{4}    &        18  &          69  &   3s 3p$^5$($^3$P$^{\sr o}$)3d $^4$D$^{\sr o}$      &             646     \\ % 3s_3p5(3Po)3d_4Do[1/2]
\ion{Ca}{5}    &         1  &           1  &                                                     &                     \\
\ion{Sc}{2}    &        38  &          85  &   3p$^6$ 3d 4f $^1$P$^{\sr o}$                      &          1\,905     \\ % 3p6_3d_4f_1Po[1]
\ion{Sc}{3}    &        25  &          45  &   7h $^2$H$^{\sr o}$                                &             454     \\ % 7h_2Ho[11/2]
\ion{Sc}{4}    &         1  &           1  &                                                     &                     \\
\ion{Ti}{2}    &        37  &         152  &   3d$^2$($^3$F)5p $^4$D$^{\sr o}$                   &          6\,173     \\ % 3d2(3F)5p_4Do[7/2]
\ion{Ti}{3}    &        33  &         206  &   3d 6f $^3$H$^{\sr o}$                             &          9\,392     \\ % 3d6f_3Ho[6]
\ion{Ti}{4}    &         1  &           1  &                                                     &                     \\
\ion{V}{1}     &         1  &           1  &   3d$^3$ 4s$^2$ a$^4$F                              &               0     \\ % 3d3_4s2_a4Fe
\ion{V}{2}     &         1  &           1  &                                                     &                     \\
\ion{Cr}{2}    &        28  &         196  &   3d$^4$($^3$G)4p x$^4$G$^{\sr o}$                  &          7\,193     \\ % 3d4(3G)4p_x4Go[11/2]
\ion{Cr}{3}    &        30  &         145  &   3d$^3$($^2$D2)4p $^3$D$^{\sr o}$                  &          4\,661     \\ % 3d3(2D2)4p_3Do[3]
\ion{Cr}{4}    &        29  &         234  &   3d$^2$($^3$P)5p $^4$P$^{\sr o}$                   &         12\,569     \\ % 3d2(3P)5p_4Po[5/2]
\ion{Cr}{5}    &         1  &           1  &                                                     &                     \\
\ion{Mn}{2}    &        25  &          97  &   3d$^4$($^5$D)4s$^2$ c$^5$D                        &             464     \\ % 3d4(5D)4s2_c5De[4]
\ion{Mn}{3}    &        30  &         175  &   3d$^4$($^3$G)4p y$^4$H$^{\sr o}$                  &          6\,292     \\ % 3d4(3G)4p_y4Ho[13/2]
\ion{Mn}{4}    &         1  &           1  &                                                     &                     \\
\ion{Fe}{1}    &        44  &         136  &   3d6($^5$D)4s 4p x$^5$F$^{\sr o}$                  &          3\,731     \\ % 3d6(5D)4s4p_x5Fo[3]
\ion{Fe}{2}    &       275  &         827  &   3d$^5$($^6$S)4p$^2$($^3$P)$^4$P                   &         89\,426     \\ % 3d5(6S)4p2(3P)4Pe[1/2]
\ion{Fe}{3}    &        69  &         607  &   3d$^5$($^4$D)6s$^3$D                              &         19\,483     \\ % 3d5(4D)6s3De[2]
\ion{Fe}{4}    &       100  &      1\,000  &   3d$^4$($^3$G)4f $^4$P$^{\sr o}$                   &        144\,005     \\ % 3d4(3G)4f_4Po[5/2]
\ion{Fe}{5}    &        47  &         191  &   3d$^3$($^4$F)4d $^5$F                             &          7\,892     \\ % 3d3(4F)4d_5Fe[3]
\ion{Fe}{6}    &        44  &         433  &   3p$^5$($^2$P)3d$^4$($^1$S) $^2$Pc$^{\sr o}$       &         27\,983     \\ % 3p5(2P)3d4(1S)_2Pco[3/2]
\ion{Fe}{7}    &        29  &         153  &   3p$^5$($^2$P)3d$^3$(b$^2$D) $^1$P$^{\sr o}$       &          3\,414     \\ % 3p5(2P)3d3(b2D)_1Po[1]
\ion{Fe}{8}    &         1  &           1  &                                                     &                     \\
\ion{Co}{2}    &        34  &         144  &   3d6($^5$D)4s 4p $^7$D$^{\sr o}$                   &          4\,112     \\ % 3d6(5D)4s4p_7Do[1]
\ion{Co}{3}    &        37  &         361  &   3d$^6$($^5$D)5p $^4$P$^{\sr o}$                   &         21\,716     \\ % 3d6(5D)5p_4Po[3/2]
\ion{Co}{4}    &        37  &         314  &   3d$^5$($^2$P)4p $^3$P$^{\sr o}$                   &         17\,220     \\ % 3d5(2P)4p_3Po[1]
\ion{Co}{5}    &        32  &         387  &   3d$^4$($^3$F)4d $^2$H                             &         27\,046     \\ % 3d4(3F)4d_2He[9/2]
\ion{Co}{6}    &        28  &         323  &   3d$^3$($^2$D)4d $^1$S                             &         19\,180     \\ % 3d3(2D)4d_1Se[0]
\ion{Co}{7}    &        31  &         319  &   3p$^5$($^2$P)d$^4$($^3$F) $^2$D$^{\sr o}$         &         18\,016     \\ % 3p5(2P)d4(3F)_2Do[3/2]
\ion{Co}{8}    &         1  &           1  &                                                     &                     \\
\ion{Ni}{2}    &        19  &          93  &   3d7($^4$F)4s 4p $^6$D$^{\sr o}$                   &          1\,639     \\ % 3d7(4F)4s4p_6Do[1/2]
\ion{Ni}{3}    &        15  &          67  &   3d$^7$($^4$F)4p $^3$D$^{\sr o}$                   &             724     \\ % 3d7(4F)4p_3Do[1]
\ion{Ni}{4}    &        36  &         200  &   3d$^6$($^3$D)4p $^2$D$^{\sr o}$                   &          8\,066     \\ % 3d6(3D)4p_2Do[5/2]
\ion{Ni}{5}    &        46  &         183  &   3d$^5$($^2$D3)4p $^3$F$^{\sr o}$                  &          6\,033     \\ % 3d5(2D3)4p_3Fo[3]
\ion{Ni}{6}    &        37  &         314  &   3d$^4$($^5$D)4d $^4$F                             &         18\,976     \\ % 3d4(5D)4d_4Fe[9/2]
\ion{Ni}{7}    &        37  &         308  &   3d$^3$($^2$D)4d $^3$P                             &         18\,364     \\ % 3d3(2D)4d_3Pe[2]
\ion{Ni}{8}    &         1  &           1  &                                                     &                     \\
\textbf{Total} & \textbf{2\,338} & \textbf{10\,605}  &                                           & \textbf{613\,214}   \\
\end{xtabular}
\flushleft
\textbf{Notes:} 
Due to a $gf$ cut (level dependent, $gf > 2 \times 10^{-3}$) only
163\,452 lines were included in the non-LTE calculations of the level
populations. 308\,846 lines were include when computing the observed
spectrum.

\clearpage

\begin{table}
\centering
\caption{Large Co model atoms used in CMFGEN calculations for the DDC10 (between 11.62\,d and 15.47\,d post explosion) and DDC25 (between 11.62\,d and 17.02\,d post explosion) models.} % based on DDC10_0p5_34
\label{tab:cmfgen_atoms_ddc_largeco_premax}
\begin{tabular}{lrrlr}
\hline
Ion & $N_\mathrm{SL}$ & $N_\mathrm{full}$ & Last level & $N_\mathrm{lines}$  \\
\hline
\ion{Co}{2}    &        81  &      1\,000  &   3d$^7$($^4$P)4f $^5$F$^{\sr o}$                   &        123\,533     \\ % 3d7(4P)4f_5Fo[4]
\ion{Co}{3}    &        85  &      1\,016  &   3d$^6$($^5$D)5f $^6$D$^{\sr o}$                   &        139\,700     \\ % 3d6(5D)5f_6Do[3/2]
\ion{Co}{4}    &        56  &      1\,000  &   3d$^5$($^2$D)5s $^1$D                             &        138\,508     \\ % 3d5(2D)5s_1De[2]
\textbf{Total} & \textbf{2\,134}  & \textbf{10\,816}  &                                          & \textbf{861\,562}   \\
\hline
\end{tabular}
\flushleft
\textbf{Notes:} 
We only report differences with respect to the model atoms shown in
Table~\ref{tab:cmfgen_atoms_ddc_early} (the total values in the last
line take into account all ions). In these calculations, we omitted
the ions C\four, O\four, Fe\six-{\sc vii}, Co\six-{\sc vii}, and
Ni\six-{\sc vii} since their populations were too low to affect the
radiative transfer.  Due to a $gf$ cut (level dependent, $gf > 2
\times 10^{-3}$) only 206\,738 lines were included in the non-LTE
calculations of the level populations. 432\,953 lines were include
when computing the observed spectrum.
\end{table}

\begin{table}
\centering
\caption{Model atoms used in CMFGEN calculations from 17.02\,d post explosion for the DDC10 model and from 18.72\,d post explosion for the DDC25 model.} % based on DDC10_0p5_38NI
\label{tab:cmfgen_atoms_ddc_postmax}
\begin{tabular}{lrrlr}
\hline
Ion & $N_\mathrm{SL}$ & $N_\mathrm{full}$ & Last level & $N_\mathrm{lines}$  \\
\hline
\ion{C}{2}     &        14  &          26  &   2s$^2$ 4d $^2$D                                   &             181     \\ % 2s2_4d_2De[5/2]
\ion{C}{3}     &        62  &         112  &   2s 8f$^1$F$^{\sr o}$                              &          1\,788     \\ % 2s8f1Fo
\ion{O}{1}     &        29  &          75  &   2s$^2$ 2p$^3$($^4$S$^{\sr o}$)5f $^3$F            &             837     \\ % 2s2_2p3(4So)5f_3Fe[2]
\ion{O}{2}     &        63  &         143  &   2s$^2$ 2p$^2$($^3$P)5p $^2$P$^{\sr o}$            &          3\,650     \\ % 2s2_2p2(3Pe)5p_2Po[3/2]
\ion{O}{3}     &        44  &          86  &   2s 2p$^2$($^4$P)3p $^3$P$^{\sr o}$                &          1\,013     \\ % 2s_2p2(4Pe)3p_3Po[0]
\ion{Ne}{2}    &        22  &          91  &   2s$^2$ 2p$^4$($^3$P)4d $^2$P                      &          2\,143     \\ % 2s2_2p4(3Pe)4d_2Pe[3/2]
\ion{Ne}{3}    &        24  &          56  &   2s$^2$ 2p$^3$($^4$S$^{\sr o}$)4p $^5$P            &             457     \\ % 2s2_2p3(4So)4p_5Pe
\ion{Mg}{2}    &        31  &          80  &   30w $^2$W                                         &          3\,863     \\ % 30w_2W
\ion{Al}{3}    &        31  &          80  &   30w $^2$W                                         &          3\,892     \\ % 30w_2W
\ion{Si}{3}    &        33  &          61  &   3s 5g$^3$G                                        &             615     \\ % 3s5g3Ge[5]
\ion{S}{2}     &        56  &         324  &   3s 3p$^3$($^5$S$^{\sr o}$)4p $^6$P                &         16\,843     \\ % 3s_3p3(5So)4p_6Pe
\ion{S}{3}     &        48  &          98  &   3s 3p$^2$($^2$D)3d $^3$P                          &          1\,633     \\ % 3s_3p2(2De)3d_3Pe
\ion{S}{4}     &        27  &          67  &   3s 3p($^3$P$^{\sr o}$)4p $^2$D                    &             761     \\ % 3s_3p(3Po)4p_2De[5/2]
\ion{Ar}{3}    &        32  &         346  &   3s$^2$ 3p$^3$($^2$D$^{\sr o}$)8s $^1$D$^{\sr o}$  &         13\,681     \\ % 3s2_3p3(2Do)8s_1Do
\ion{Ti}{2}    &        61  &      1\,000  &   3d$^2$($^3$F)9p $^4$F$^{\sr o}$                   &        185\,756     \\ % 3d2(3F)9p_4Fo[7/2]
\ion{Cr}{2}    &        28  &         196  &   3d$^4$($^3$G)4p x$^4$G$^{\sr o}$                  &          8\,249     \\ % 3d4(3G)4p_x4Go[11/2]
\ion{Fe}{3}    &        83  &         698  &   3d$^5$($^2$H)4d 1K                                &         73\,419     \\ % 3d5(2H)4d_1Ke[7]
\ion{Co}{2}    &       109  &         948  &   3d$^7$($^4$P)6s $^3$P                             &        189\,440     \\ % 3d7(4P)6s_3Pe[1]
\ion{Co}{3}    &        85  &      1\,016  &   3d$^6$($^5$D)5f $^6$D$^{\sr o}$                   &        139\,700     \\ % 3d6(5D)5f_6Do[3/2]
\ion{Co}{4}    &        56  &      1\,000  &   3d$^5$($^2$D)5s $^1$D                             &        139\,240     \\ % 3d5(2D)5s_1De[2]
\ion{Ni}{2}    &        59  &      1\,000  &   3d$^8$($^3$F)7f $^4$I$^{\sr o}$                   &        103\,224     \\ % 3d8(3F)7f_4Io[9/2]
\ion{Ni}{3}    &        47  &      1\,000  &   3d$^7$($^2$D)4d $^3$Sb                            &        132\,677     \\ % 3d7(2D)4d_3Sbe[1]
\ion{Ni}{4}    &        54  &      1\,000  &   3d$^6$($^5$D)6p $^6$F$^{\sr o}$                   &        145\,527     \\ % 3d6(5D)6p_6Fo[11/2]
\textbf{Total} & \textbf{2\,351} & \textbf{14\,434} &                                            & \textbf{1\,539\,740}\\
\hline
\end{tabular}
\flushleft
\textbf{Notes:} 
We only report differences with respect to the model atoms shown in
Table~\ref{tab:cmfgen_atoms_ddc_early} (the total values in the last
line take into account all ions).  Due to a $gf$ cut (level dependent,
$gf > 10^{-4}$) only 586\,635 lines were included in the non-LTE
calculations of the level populations. 772\,836 lines were include
when computing the observed spectrum.
\end{table}

\begin{table}
\centering
\caption{Large Co model atoms used in CMFGEN calculations for the DDC10 and DDC25 models from 33.15\,d post explosion onward.} % based on DDC10_0p5_45NIH
\label{tab:cmfgen_atoms_ddc_largeco_postmax}
\begin{tabular}{lrrlr}
\hline
Ion & $N_\mathrm{SL}$ & $N_\mathrm{full}$ & Last level & $N_\mathrm{lines}$  \\
\hline
\ion{Co}{2}    &       136  &      2\,747  &   3d$^7$($^2$D)6p $^3$P$^{\sr o}$                   &     1\,186\,076     \\ % 3d7(2D)6p_3Po[1]
\ion{Co}{3}    &       124  &      3\,917  &   3d$^6$($^3$D)6d $^4$P                             &     1\,357\,405     \\ % 3d6(3D)6d_4Pe[3/2]
\textbf{Total} & \textbf{2\,292}  & \textbf{18\,373} &                                           & \textbf{3\,713\,110}\\
\hline
\end{tabular}
\flushleft
\textbf{Notes:} 
We only report differences with respect to the model atoms shown in
Table~\ref{tab:cmfgen_atoms_ddc_postmax} (the total values in the last
line take into account all ions).  Due to a $gf$ cut (level dependent,
$gf > 10^{-4}$), only 1\,277\,321 lines were included in the non-LTE
calculations of the level populations. We note that 1\,860\,140 lines were included
when computing the observed spectrum.
\end{table}

\begin{table}
\centering
\caption{Reduced Fe\four, Co\four, and Ni\four\ model atoms used in CMFGEN calculations for the DDC10 and DDC25 models from 40.11\,d post explosion onward.} % based on DDC10_0p5_47NIH
\label{tab:cmfgen_atoms_ddc_reducednicofe4}
\begin{tabular}{lrrlr}
\hline
Ion & $N_\mathrm{SL}$ & $N_\mathrm{full}$ & Last level & $N_\mathrm{lines}$  \\
\hline
\ion{Fe}{4}    &        35  &         176  &   3d$^4$($^3$G)4p $^4$H$^{\sr o}$                   &          6\,595     \\ % 3d4(3G)4p_4Ho[13/2]
\ion{Co}{4}    &        37  &         314  &   3d$^5$($^2$P)4p $^3$P$^{\sr o}$                   &         17\,952     \\ % 3d5(2P)4p_3Po[1]
\ion{Ni}{4}    &        28  &         254  &   3d$^6$($^1$G1)4p $^2$G$^{\sr o}$                  &         12\,512     \\ % 3d6(1G1)4p_2Go[7/2]
\textbf{Total} & \textbf{2\,182} & \textbf{16\,117} &                                            & \textbf{3\,321\,397}\\
\hline
\end{tabular}
\flushleft
\textbf{Notes:} 
We only report differences with respect to the model atoms shown in
Table~\ref{tab:cmfgen_atoms_ddc_postmax} (the total values in the last
line take into account all ions).  Due to a $gf$ cut (level dependent,
$gf > 10^{-4}$), only 1\,118\,403 lines were included in the non-LTE
calculations of the level populations. We note that 1\,663\,922 lines were include
when computing the observed spectrum.
\end{table}

\subsection{SEDONA}\label{sect:atomdata_sedona}

SEDONA used the Kurucz CD 1 line list to compute the bound-bound opacities. 

\begin{table}
\centering
\caption{Bound-bound atomic data information from the Kurucz CD 1 line list for the toy06 and toy01 models computed with SEDONA}\label{tab:sedonadatatoy06}
\begin{tabular}{cccr}
\hline
Element $Z$ & Element $A$ & $N_\mathrm{stages}$ & $N_\mathrm{lines}$ \\  \hline
 28 & 56 & 6 & 7\,609\,586  \\ 
 28 & 58 & 6 & 7\,609\,586  \\ 
 27 & 56 & 6 & 8\,024\,034 \\
 26 & 56 & 6 & 6\,620\,297 \\
 20 & 40 & 6 &    632\,282 \\
 16 & 32 & 6 &      2\,289 \\
 14 & 28 & 6 &      8\,709 \\
\textbf{Total} &  &  & \textbf{22\,897\,197}\\
\hline
\end{tabular}
\end{table}

\begin{table}
\centering
\caption{Bound-bound atomic data information from the Kurucz CD 1 loneliest for the DDC10 model computed with SEDONA}\label{tab:sedonadataddc10}
\begin{tabular}{cccr}
\hline
Element $Z$ & Element $A$ & $N_\mathrm{stages}$ & $N_\mathrm{lines}$ \\  \hline
 28 & 56 & 5 & 5\,543\,966  \\ 
 28 & 58 & 5 & 5\,543\,966  \\ 
 27 & 57 & 5 & 7\,066\,365 \\
 27 & 56 & 5 & 7\,066\,365 \\
 26 & 56 & 5 & 6\,175\,428 \\
 24 & 52 & 5 & 3\,010\,793 \\
 24 & 48 & 5 & 3\,010\,793 \\
 23 & 48 & 5 & 2\,226\,269 \\
 22 & 48 & 5 &    939\,560 \\
 22 & 44 & 5 &    939\,560 \\
 20 & 40 & 4 &    101\,996 \\
 16 & 32 & 5 &      2\,142 \\
 14 & 28 & 5 &      8\,705 \\
 12 & 24 & 3 &      3\,127 \\
 10 & 20 & 3 &     11\,428 \\
  8 & 16 & 6 &      8\,062 \\
  6 & 12 & 6 &      8\,139 \\
\textbf{Total} &  &  & \textbf{25\,105\,980}\\
\hline
\end{tabular}
\end{table}

\subsection{TARDIS}\label{sect:atomdata_tardis}

TARDIS used the Kurucz CD 23 line list to compute the bound-bound
opacities. The atomic data file was generated with the \textsc{Carsus}
package on August 24, 2017, named \verb|kurucz_cd23_chianti_H_He.h5|
and signed with UUID1 \verb|6f7b09e887a311e7a06b246e96350010| and MD5
\verb|864f1753714343c41f99cb065710cace|.

\hfill

Table~\ref{tab:tardisdata} tabulates the total number of levels
($N_\mathrm{levels}$), meta-stable levels ($N_\mathrm{meta}$), and
lines ($N_\mathrm{lines}$) for the atoms used in the four
models. Atoms present in the atomic data file but not used by the
models were not listed.

\clearpage

\centering
\xentrystretch{-.1}
\topcaption{Bound-bound atomic data information from the Kurucz CD 23 line list for all models computed with TARDIS.}
\label{tab:tardisdata}
\tablefirsthead{\toprule Ion & $N_\mathrm{levels}$ & $N_\mathrm{meta}$ & $N_\mathrm{lines}$ \\ \midrule}
\tablehead{
\multicolumn{4}{l}{\footnotesize \textbf{Table~\ref*{tab:tardisdata}.} continued.} \\
\multicolumn{4}{l}{} \\
\toprule
Ion & $N_\mathrm{levels}$ & $N_\mathrm{meta}$ & $N_\mathrm{lines}$ \\ 
\midrule
}
\tabletail{\midrule}
\tablelasttail{\bottomrule}
\begin{xtabular}{lrrr}
     \ion{C}{1} &             833 &            603 &             3249 \\
     \ion{C}{2} &              86 &              6 &              374 \\
     \ion{C}{3} &              81 &              6 &              388 \\
     \ion{C}{4} &              36 &              1 &              192 \\
     \ion{O}{1} &             150 &              9 &              854 \\
     \ion{O}{2} &             173 &              6 &             1374 \\
     \ion{O}{3} &             141 &              9 &              766 \\
     \ion{O}{4} &             146 &             19 &              465 \\
     \ion{O}{5} &              97 &              5 &              459 \\
    \ion{Ne}{1} &             284 &             23 &             2422 \\
    \ion{Ne}{2} &             283 &              9 &             3468 \\
    \ion{Ne}{3} &              64 &              9 &              269 \\
    \ion{Ne}{4} &              99 &             12 &              340 \\
    \ion{Ne}{5} &              64 &             13 &              164 \\
    \ion{Na}{1} &              58 &              1 &              334 \\
    \ion{Na}{2} &              35 &              3 &              171 \\
    \ion{Na}{3} &              69 &              4 &              353 \\
    \ion{Na}{4} &              46 &              5 &              110 \\
    \ion{Na}{5} &              71 &             10 &              187 \\
    \ion{Mg}{1} &             552 &            366 &             1580 \\
    \ion{Mg}{2} &              75 &              2 &              510 \\
    \ion{Mg}{3} &              93 &              4 &              704 \\
    \ion{Mg}{4} &              54 &              5 &              169 \\
    \ion{Mg}{5} &              53 &              5 &              132 \\
    \ion{Al}{1} &             273 &            160 &              482 \\
    \ion{Al}{2} &             197 &             10 &             2602 \\
    \ion{Al}{3} &              58 &              1 &              342 \\
    \ion{Al}{4} &              31 &              3 &              142 \\
    \ion{Al}{5} &              56 &              9 &               77 \\
    \ion{Si}{1} &             558 &            230 &             3856 \\
    \ion{Si}{2} &             100 &             13 &              567 \\
    \ion{Si}{3} &             169 &             10 &             1248 \\
    \ion{Si}{4} &              52 &              1 &              307 \\
    \ion{Si}{5} &              35 &              3 &              125 \\
     \ion{S}{1} &             153 &             12 &              727 \\
     \ion{S}{2} &              85 &              7 &              500 \\
     \ion{S}{3} &              58 &             22 &              170 \\
     \ion{S}{4} &              28 &              5 &               50 \\
     \ion{S}{5} &              19 &              5 &               41 \\
    \ion{Cl}{1} &             229 &             23 &             2542 \\
    \ion{Cl}{2} &             128 &             21 &              973 \\
    \ion{Cl}{3} &              78 &             13 &              431 \\
    \ion{Cl}{4} &              33 &              5 &              121 \\
    \ion{Cl}{5} &              27 &              5 &               43 \\
    \ion{Ar}{1} &             215 &              8 &             2397 \\
    \ion{Ar}{2} &             314 &             26 &             4567 \\
    \ion{Ar}{3} &              96 &             16 &              655 \\
    \ion{Ar}{4} &              39 &              7 &              104 \\
    \ion{Ar}{5} &              22 &              6 &               49 \\
     \ion{K}{1} &              94 &             15 &              575 \\
     \ion{K}{2} &              22 &              4 &               66 \\
     \ion{K}{3} &              40 &              5 &              192 \\
     \ion{K}{4} &              24 &              5 &               57 \\
     \ion{K}{5} &              33 &              9 &               75 \\
    \ion{Ca}{1} &             198 &              5 &             2906 \\
    \ion{Ca}{2} &              93 &              3 &              752 \\
    \ion{Ca}{3} &             150 &             13 &             1766 \\
    \ion{Ca}{4} &              70 &             26 &              122 \\
    \ion{Ca}{5} &              39 &              5 &               91 \\
    \ion{Sc}{1} &             272 &             11 &             4221 \\
    \ion{Sc}{2} &             168 &             14 &             2215 \\
    \ion{Sc}{3} &              43 &              3 &              217 \\
    \ion{Sc}{4} &             127 &             13 &              953 \\
    \ion{Sc}{5} &              22 &              6 &               29 \\
    \ion{Ti}{1} &             441 &             38 &             8771 \\
    \ion{Ti}{2} &             204 &             37 &             2597 \\
    \ion{Ti}{3} &             199 &             13 &             2289 \\
    \ion{Ti}{4} &              39 &              3 &              139 \\
    \ion{Ti}{5} &              51 &             13 &              331 \\
     \ion{V}{1} &             502 &             60 &             6995 \\
     \ion{V}{2} &             323 &             75 &             4545 \\
     \ion{V}{3} &             299 &             35 &             5304 \\
     \ion{V}{4} &              98 &             13 &              995 \\
     \ion{V}{5} &              64 &              3 &              335 \\
    \ion{Cr}{1} &             394 &             83 &             4172 \\
    \ion{Cr}{2} &             733 &             97 &            17224 \\
    \ion{Cr}{3} &             214 &             71 &             2122 \\
    \ion{Cr}{4} &             154 &             35 &             1717 \\
    \ion{Cr}{5} &              46 &             13 &              220 \\
    \ion{Mn}{1} &             322 &             52 &             3023 \\
    \ion{Mn}{2} &             569 &             94 &             8362 \\
    \ion{Mn}{3} &             391 &             95 &             5848 \\
    \ion{Mn}{4} &             103 &             40 &              677 \\
    \ion{Mn}{5} &              84 &             35 &              602 \\
    \ion{Fe}{1} &             848 &             68 &            22905 \\
    \ion{Fe}{2} &             796 &             85 &            21753 \\
    \ion{Fe}{3} &             566 &             97 &             9860 \\
    \ion{Fe}{4} &             276 &             99 &             3559 \\
    \ion{Fe}{5} &             180 &             71 &             1865 \\
    \ion{Co}{1} &             317 &             46 &             5298 \\
    \ion{Co}{2} &             256 &             47 &             2853 \\
    \ion{Co}{3} &             213 &             58 &             2247 \\
    \ion{Co}{4} &             296 &             96 &             4092 \\
    \ion{Co}{5} &             267 &             94 &             3542 \\
    \ion{Ni}{1} &             180 &             17 &             2671 \\
    \ion{Ni}{2} &             717 &             25 &            17150 \\
    \ion{Ni}{3} &             344 &             47 &             5456 \\
    \ion{Ni}{4} &             235 &             70 &             2712 \\
    \ion{Ni}{5} &             323 &            101 &             4733 \\
 \textbf{Total} &  \textbf{19135} &  \textbf{3819} &  \textbf{243353} \\
\end{xtabular}

\clearpage

\end{appendix}

\end{document}